\documentclass[%
 reprint,
 amsmath,amssymb,
 aps,
floatfix,
]{revtex4-1}

\usepackage{graphicx}% Include figure files
\usepackage{dcolumn}% Align table columns on decimal point
\usepackage{bm}% bold math
\usepackage{bbm}
\usepackage{macros}

\newcommand{\lr}[1]{\left( #1 \right)}

\begin{document}
\sloppy

%\preprint{APS/123-QED}

\title{Reduced critical slowing down for statistical physics
  simulations }

\author{Kurt Langfeld}
\affiliation{
  School of Mathematics, University of Leeds, Leeds, LS2 9JT, UK
}

\author{Pavel Buividovich}
\author{P.E.L Rakow}
\author {James Roscoe}

\affiliation{
 Department of Mathematical Sciences, University of Liverpool, Liverpool,
L69 7ZX, UK
}

\date{\today}

\begin{abstract}
Wang-Landau simulations offer the possibility to integrate explicitly
over a collective coordinate and stochastically over the remainder of
configuration space. We propose to choose the so-called ``slow mode'',
which is responsible for large autocorrelation times and thus critical
slowing down, for collective integration. We study this proposal for
the Ising model and the linear-log-relaxation (LLR) method as 
simulation algorithm. We firstly demonstrate super critical slowing
down in a phase with spontaneously broken symmetry and for the
heatbath algorithms, for which autocorrelation times grow
exponentially with system size. By contrast,  using the magnetisation as
collective coordinate, we present evidence that super critical slowing down
is absent. We still observe a polynomial increase of
the autocorrelation time with volume (critical slowing down), which is
however reduced by orders of magnitude when compared to local
update techniques. 
\end{abstract}

\maketitle

% ----------------------------------------------------------------------
\section{Introduction}

Stochastic simulations of lattice theories combined with modern
computer resources have rapidly evolved to an exceptional theoretical
framework enlightening research areas such as Quantum Field
Theory~\cite{Rothe} and Statistical
Physics~\cite{binder_heermann_montecarlo}. Markov Chain Monte Carlo (MCMC)
simulations in conjunction with a local update of
the degrees of freedom  are ubiquitous in the quiver of
possibilities.

\medskip
In MCMC simulations, a bunch of local updates - usually called MC sweep -
result into a new configuration of degrees of freedom on the
lattice. The simulations generates sequentially a string of lattice
configurations. Under the Markov assumption, any configuration only
depends on its predecessor. Objects of interests  are expectation
values. By virtue of the law of large numbers~\cite{Dekking}, those
can be estimated using the $N$ configurations of the Markov
set:
$$
\langle A \rangle \; \approx \; \frac{1}{N} \sum _{i=1}^N A_i \; .
$$
The price to pay for a finite reach $N$ is that the above estimator is
afflicted by a statistical error $\epsilon_A$, which scales like
$1/\sqrt{N}$ under the Markov assumption (and assuming that the variance
of $A$ exists). 

\medskip
In practical Monte-Carlo simulations, configurations are correlated
over a characteristic number of Monte-Carlo updates $t \approx  \tau
$, which is called 
autocorrelation time (we 
will give a proper definition below). An immediate impact
is that the statistical error now scales like $\sqrt{\tau / N}$. Large
autocorrelations times severely limit the usefulness of simulations
at moderate computational costs, and a good deal of algorithmic
research has been devoted to simulation methods with small
autocorrelations.

\medskip
The autocorrelation time depends on the simulation algorithm, the
parameters of the simulated theory and the system size, say volume
$V$, which could be the number of lattice sites. Of particular
interest for many applications is a parameter regime that leaves the
lattice degrees of freedom correlated over a typical spatial scale
$\xi $ (correlation length). In Solid State Physics, $\xi $ diverges
at a second order phase transition. In quantum physics simulations
$1/\xi $ acts a regulator for the inherent divergencies of the
underpinning quantum field theory, and the limit $\xi \to \infty $
is of crucial importance to extract physics relevant information from
those computer simulations. Generating independent Markov ensembles in
the case that degrees of freedom are correlated over many sites is a
challenge for any algorithm and in particular for the important class of {\it local
  update algorithms }. This challenge is reflected by the
monotonically increasing function $\tau (\xi )$ which describes the
connection between correlation 
length $\xi $ and the autocorrelation time $\tau $. On a finite
lattice, say with an extent $L$, spatial correlations are limited by
$L$, leaving us with: $\tau = \tau (L)$. We will distinguish between a
power-law and an exponential relation: 
\bea 
\tau (L) &\propto & L ^{z} \; , \; \; \; \hbox{(critical slowing down) }
\nonumber \\
\hbo \tau (L) &\propto & \e^{ m\, L } , \; \; \; \hbox{(super critical
  slowing down). }
\nonumber 
\ena 
Because of the connection between autocorrelation time $\tau $ and
statistical error $\epsilon $, theories in the parameter regime
afflicted by super critical slowing down can only be simulated for small or
moderate lattice
sizes $L$, and extrapolation to large $L$ might or might not be
possible.

\medskip
Over many decades, research has been analysing the combination of
theories and algorithms studying autocorrelations times for
particular observables. For Markov chain simulation that satisfy
detailed balance, large autocorrelations times are traced back to low
eigenvalues of the transition matrix~\cite{Schaefer:2010hu}. The latter
paper offers a detailed study for lattice QCD and the important Hybrid
Monte Carlo approach~\cite{Duane:1987de}. In theories that admit a
characterisation of configurations by topology, such as QCD and CP(N)
models, critical slowing down is often related to slowly-evolving
topological modes~\cite{Bonati:2017woi,Brower:2003yx}. More generally, modes with
slowest de-correlation typically correspond to long-wavelength modes of
physical fields. For a free scalar field theory, a combination of
particular order of updating the fields and tuning of stochastic
overrelation {\it can} significantly reduce critical slowing
down~\cite{HORVATH1998367}. Albeit this is per se an interesting
finding, we here do not consider algorithms that need significant fine
tuning for reducing autocorrelations. 

\medskip
To alleviate the ``slow mode relaxation'' issue, multigrid methods
have been proposed already in the late eighties~\cite{Stoll_1989}.
For specific models, targeted solutions can be found that either
eliminate critical slowing down or strongly reduce it. Those attempts are a based on a
reformulation, and simulations include non-local updates. For the CP(N-1)
model, which is plagued by the slow mode issue due to topological
sectors, a complete absence of critical slowing down was reported
in~\cite{Wolff:2010qz} for two dimensions. {\it Cluster update
  algorithms}~\cite{Swendsen:1987ce,Wolff89} generically possess a
small dynamical critical exponent $z$ and thus provide a practical
solution to the critical slowing down issue. Whenever a model allows a
cluster reformulation, the performance cluster algorithms are hardly
outperformed by any other approach and hence are the preferred
simulation method. 

\medskip
Lattice theories that show {\it spontaneous symmetry breaking} in the
infinite volume limit are particularly prone to {\it super critical
  slowing down} when simulated in the broken phase. Let $\phi_x$  be
the fields of such a theory with partition function
$$
Z(\beta ) \; = \; \int {\cal D}\phi \; \exp \{ \beta \, S(\phi)\} \; ,
$$
and $M(\phi )$
the order parameter. For any finite lattice size, the symmetry
implies that the expectation value of the order parameter, i.e.,
$\langle M \rangle $ vanishes. In the broken phase, stochastically
``important'' configurations cluster in domains with $M(\phi )
\not=0$~\cite{binder1981}, and $\langle M \rangle $ vanishes upon
averaging over these 
relevant domains. Local update algorithms usually fail to induce
transitions between these domains leading to super critical slowing
down. Yang-Mills
theories with a gauge group $SU(N\ge3)$ fall into this important
class of models~\cite{HollandWiese}.  Gauge symmetry prevents the
definition of meaningful (gauge invariant) clusters and corresponding
non-local update algorithms. We are hence turning to other more
conventional  simulation  techniques.

\medskip
A promising class of such algorithms are multi-canonical
algorithms~\cite{Billoire_1993} and Wang-Landau
techniques~\cite{WangLandau2001,Wang2001DeterminingTD}. Although the
algorithmic differences and similarities between both methods have been
studied in the literature (see e.g.~\cite{junghans2014}), both employ
reweighing techniques with respect to a marginal distribution, which is at the
heart of solving the issue of super critical slowing down. This has
been firstly demonstrated by Torrie and Valleau~\cite{TORRIE1977187}
in a thermodynamics setting and later by Berg, Hansmann and Neuhaus for the ising
model in~\cite{berg1993}. 

\medskip
At the root of super critical slowing down  is the double-peak
marginal distribution $P(M)$ of the order parameter, say the
magnetisation $M$. Rather than leave it to importance sampling to
transition between the two equally important phases, we calculate the
partition function by integrating {\it  explicitly } over the order
parameter $M$ and stochastically over the remainder of the configuration
space. To this aim, we exploit the identity
\bea
Z(\beta ) &=& \int dm \; \rho (m) \; ,
\nonumber \\
\rho (m) &=& \int  {\cal D}\phi \; \delta \Bigl( m - M(\phi
) \Bigr) \exp \{ \beta \, S(\phi)\} \; ,
\nonumber
\ena
where $\delta $ is the Dirac $\delta -$function. Thereby, $\rho $ is
called the density-of-states. Density-of-states techniques have seen
remarkable successes over the last decade ranging from a study of the
QCD phase diagram at significant baryon chemical
potentials~\cite{Fodor:2007vv}, a recent study of the topological
density in pure Yang-Mills theories~\cite{Borsanyi:2021gqg} and the
first proof of concept of solving a strong sign-problem using the
$Z_3$ theory~\cite{Langfeld:2014nta}.

\medskip
Key to the success of the density-of-states techniques is a robust
method to estimate the density-of-states $\rho $ including control
over its stochastic errors. In this paper, we explore the {\it
  Linear-Log-Relaxation} (LLR)
method~\cite{Langfeld:2012ah,Langfeld:2015fua,Langfeld:2016kty}, which
belongs to the class of the Wang-Landau techniques. The LLR method is
based upon a systematical expansion of the marginal distribution $\rho
(m)$ in a given $m$-interval leading to a stochastic non-linear
equations for the expansion parameters (see below for details). In its
lowest order, the LLR approach has similarities with the
``multi-magnetic ensemble'' method by Berg, Hansmann and
Neuhaus~\cite{berg1993}.  The LLR-approach is also markedly
different: it confines the MC simulation part to a window of size $2
\delta $ around a given value of the magnetisation $m_0$, which is a
non-local constraint. We will be interested in the limit $\delta \to
0$. 

\medskip
In this paper, we offer a systematic and large scale study of the
phenomenon of critical slowing down using the LLR method. Since we are
interested in simulation methods, which can applied universally to a
wide range of lattice models, we benchmark our findings against those
from a heatbath approach rather than a cluster algorithm, which would
be anyhow the method of choice if applicable. We find evidence that
{\it super critical slowing down } is absent (in line with the
findings from a multi-canonical simulation~\cite{berg1993}). We still
find a correlation length that increases polynomial with the
volume. We observe, however, that those correlations are strongly
suppressed even at criticality.

%----------------------------------------------------------------------
\section{Understanding critical slowing down}

\subsection{Accessing autocorrelations}

The well-studied Ising model in a finite volume also serves here to illustrate the
breakdown of {\it importance sampling } due to a failure of sampling the
configuration space within an acceptable amount of computational
resources. The purpose of this section is to quantify this breakdown
for the popular Markov-Chain Monte-Carlo (MCMC) approach. We are
particularly interested in the parameter dependence of failure,
foremost its dependence on the system size. All numerical illustrations
of this section are carried out using shockingly small lattice
sizes. This illustrates the severeness of the issue: These small sizes
are mandatory because  of the rapid breakdown of ergodicity at even
moderate lattice sizes.

\medskip
Protagonists are the Ising spins $s_x = \pm 1$ associated with each
lattice site $x$ of the lattice of size $V=L \times L $. We use
periodic boundary with periodic boundary conditions~\cite{Ising1925}
throughout the paper. Partition function $Z$ and action $S$ are
given by
\be
Z = \sum_{\{s_x\}} \exp \{ \beta S \} \; , \hbo
S = \sum _{\langle xy\rangle } s_x s_y \; ,
\label{eq:1}
\en
where the sum in the action extends over all nearest neighbours $x$
and $y$. Results for autocorrelations will depend on the
algorithm. We therefore present details of the simulation here. We are
employing the standard  {\it heatbath algorithm} as benchmark:

\begin{enumerate}
\item Choose a site $x$ of the lattice at random, and calculate the
  sum over the neighbouring spins:
  $$
  b_x \; = \; \sum _{y \in \langle xy \rangle } s_y \; .
  $$
\item Define
  $$
p_x \; = \; \frac{1}{1 + \exp \{ - 2 \beta \, b_x\} } \; ,
$$
and choose $s_x=1$ with probability $p_x$ and set $s_x=-1$ otherwise.
\item Repeat both steps 1-2 above $V$ times to complete one lattice
  sweep.

\item The spin configuration $\{s_x\}_k$ after $k$ sweeps is
  considered as part of a chain of configurations labeled by the
  Monte-Carlo time $k=1 \ldots N$. Define a sequence of random numbers
  for an observable $f(\{s_x\})$ by
  $$
f_1 \to f_2 \to \ldots f_N \; , \hbo f_i \; = \; f\Bigl(\{s_x\}_i
\Bigr) .
$$

\item Obtain estimators for observables by
  $$
f \; := \; \frac{1}{N} \sum _{i=1}^N
f_i \; .
$$

\item Repeating steps 1-5 many times defines a random process for $f$
  itself. We denote the corresponding average by $[f]$. Note that $[f]$
  is hence independent of, e.g., the random numbers used for a
  particular run, but does depend on $N$. Approximate
  $$
\langle f \rangle \; \approx \; [f] \; .
  $$

\end{enumerate}

\begin{figure}
  \includegraphics[height=8cm]{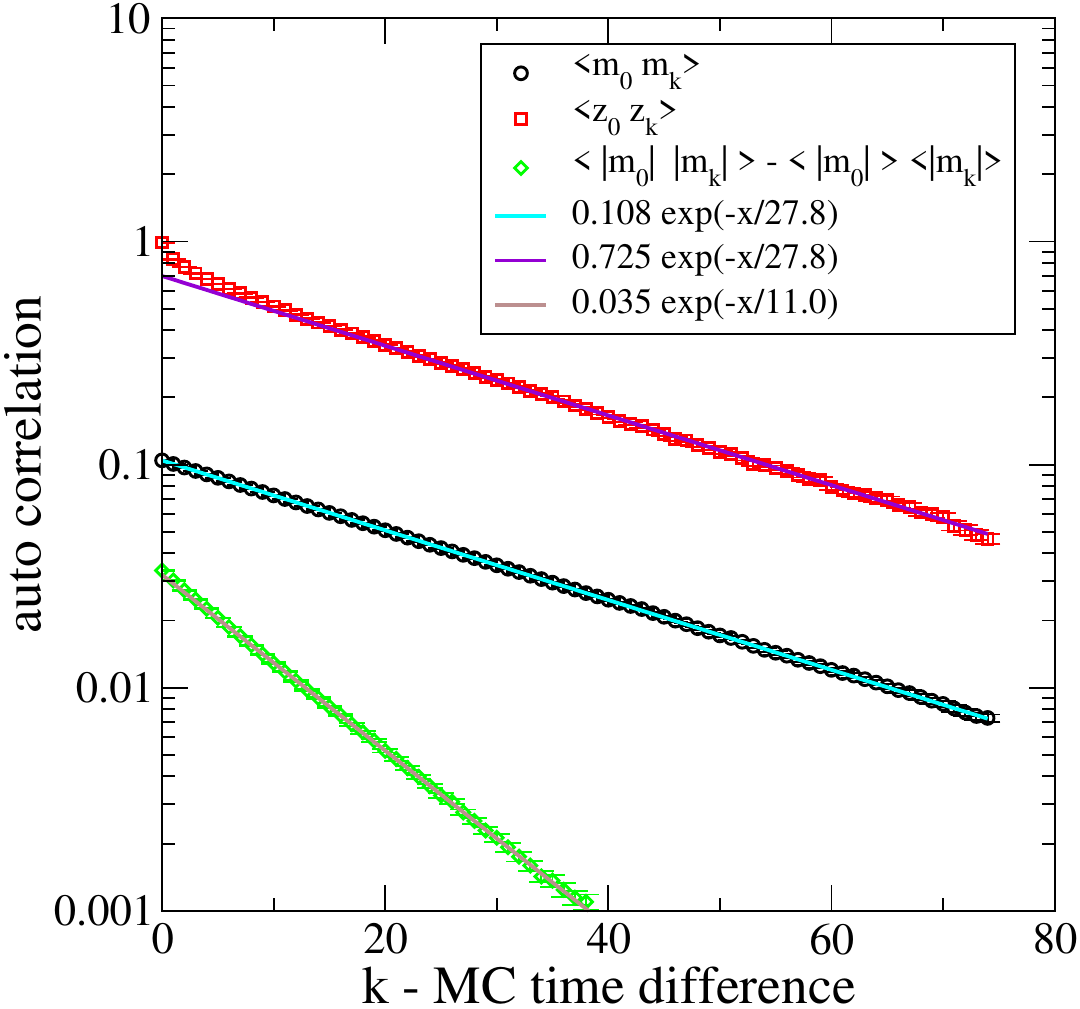}
  \caption{\label{fig:1}
    Autocorrelation functions for a
    $12\times 12$ Ising model at $\beta =0.35$  as a function of the MC time
    difference $k$ (see~\ref{eq:9}).}
\end{figure}
\medskip
A variable of particular interest is the {\it magnetisation per spin}
$$
\langle m \rangle \; = \; \left \langle \frac{1}{V} \sum _x^V s_x
\right\rangle \;  = \; \langle s_x \rangle \; ,
$$
which does not depend on the site $x$ due to translation invariance.
The corresponding elements of the chain of random variables are given
by
\be
m_i \; = \; \frac{1}{V} \sum _{x=1}^V s_x^{(i)} ,
\label{eq:3}
\en
where $s_x^{(i)}$ is the spin at site $x$ of the configuration
$\{s_x\}_i$.

\medskip
By the law of large numbers, we find
$$
\langle m \rangle \; = \; \lim _{N\to \infty }  [m](N) \; .
$$
Any stochastic simulation, however, resorts to a finite length $N$ of
the chain, and the central question is to what extend
is the approximation
\be
\langle m \rangle \approx [m]
\label{eq:approx}
\en
valid?

\medskip
To avoid a cluttering of notation, we preemptively use a result of
the next subsection. By virtue of  a symmetry argument, we have
$$
\langle s_x \rangle \; = \; 0 \; , \hbo [m](N) \; = \; 0 \; , \; \; \;
\forall N \; .
$$
As usual, the error for the approximation (\ref{eq:approx}) is given
by the standard deviation
\be
\epsilon ^2 \; = \; \left[ m^2 \right] \; - \; [ m ]^2
\; = \;  \left[ m^2 \right] .
\label{eq:7}
\en
We find
\be
\epsilon ^2 \; = \; \left[ \left( \sum _{i=1}^N \sum _{\ell=1}^N  m_i
  \right) \right] \; = \; \frac{1}{N^2} \sum _{i=1}^N \sum _{\ell=1}^N \left[ m_i \,
  m_\ell \right]\; .
\label{eq:8}
\en
Apparently, the latter equation depends how the random variable $m_i$
is correlated to the variable $m_\ell$, and the average $m_i m_\ell$ is
called {\it autocorrelation}. A key assumption here is that this
correlation decreases exponentially with the distance $\vert k \vert
$ between the positions in the chain:
\bea
\left[ m_i \,  m_\ell \right] &=&m_0^2 \; \exp \left\{ -
  \frac{ k }{\tau } \right\} \; , \; \; \; k = \vert i-\ell \vert 
\label{eq:9} \\
m_0^2 &:=& \left[ m_i^2 \right] \; ,
\nonumber
\ena
where $\tau $ is called {\it autocorrelation } time. This is
expected  to be the case for large separations $k$. A rather stark
assumption is that the exponential behaviour dominates the double sum
in (\ref{eq:8}). This assumption only can be justified afterwards in the
numerical experiment but it seems to be the case for the parameter
range explored in this paper. Inserting
(\ref{eq:9}) into (\ref{eq:8}), the double sum can be performed
analytically:
\bea
\epsilon ^2 &=&
\frac{m_0^2  }{N^2} \sum_{\ell=1}^N \sum _{i=1}^N
a^{\vert i-\ell\vert}
\nonumber \\
&=& \frac{m_0^2  }{N} \, \frac{1+a}{1-a} -
\frac{2a \, m_0^2 }{N^2 (1-a)^2} \, \left(1-a^N\right) \; ,
\label{eq:10} \\
a &=& \exp \{ -1/\tau \} \; .
\label{eq:11}
\ena
For a moderately sized autocorrelation time, we might find ourselves in a
situation where we have $1 \ll \tau \ll N $. Expanding (\ref{eq:10}) yields for
this case:
\be
\epsilon ^2 \; = \; \frac{2 m_0^2 \,  \tau }{N}
\; + \; {\cal O} \left(  \frac{\tau ^2 }{N^2 } \right) \; .
\label{eq:12}
\en
This the famous $1/\sqrt{N}$ law of MCMC simulations taking into
account an autocorrelation time $\tau \gg 1 $.

\medskip
In  case that the autocorrelation time is exceedingly large, we might face the ordering
$ 1 \ll N \ll \tau $. Expanding (\ref{eq:10}) for this scenario yields
an entirely different picture:
\be
\epsilon ^2 \; = \; m_0^2 \; \left[ 1 \, - \,
\frac{N}{3 \tau } \; + \; {\cal O} \left( \frac{1}{N \tau},
  \frac{N^2}{\tau ^2} \right) \, \right] \; .
\label{eq:13}
\en
In this case, the error is of order one, and we cannot expect that
(\ref{eq:approx}) yields a meaningful approximation. Note, however,
that equation (\ref{eq:13}) still can provide information on the
(large) autocorrelation time by virtue of the correction to the
leading term even if $N \sim \tau $.

\subsection{Symmetry breaking  and ergodicity}

\begin{figure*}
  \includegraphics[height=8cm]{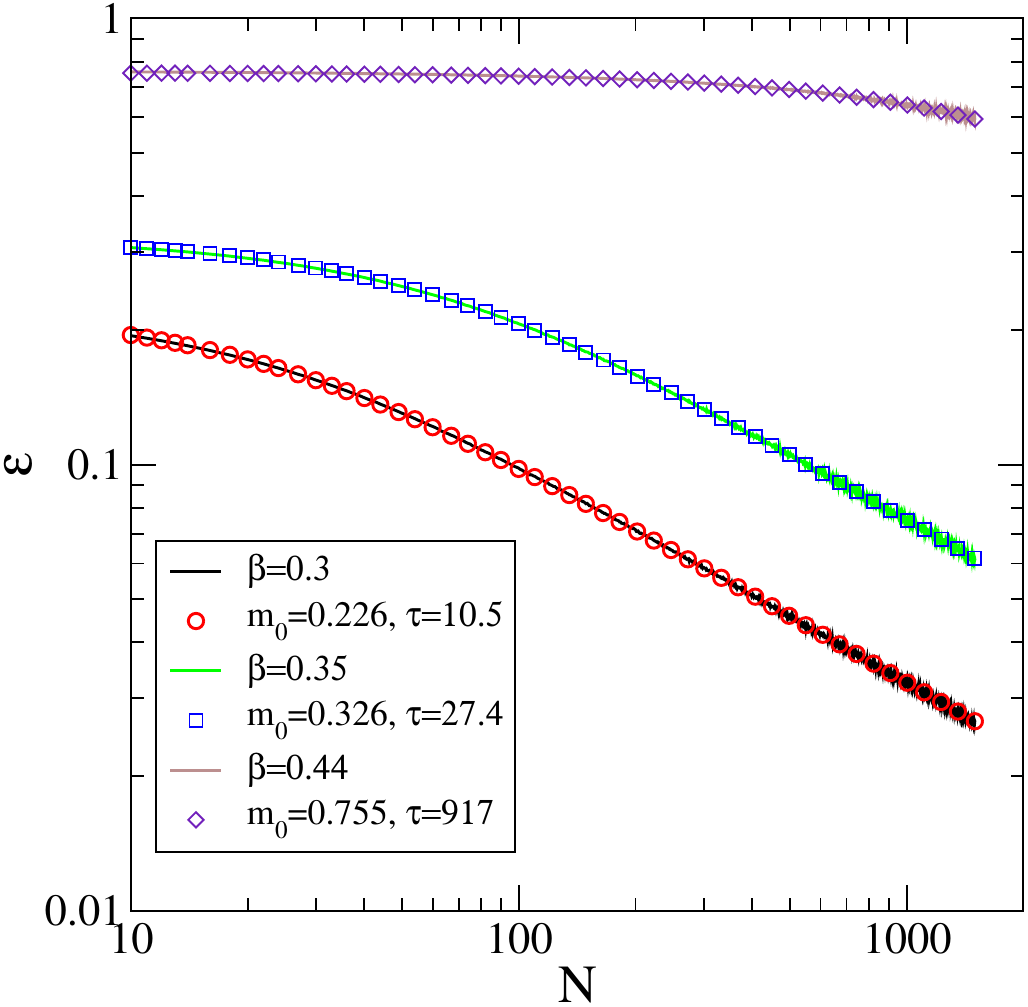} \hspace{0.5cm}
  \includegraphics[height=8cm]{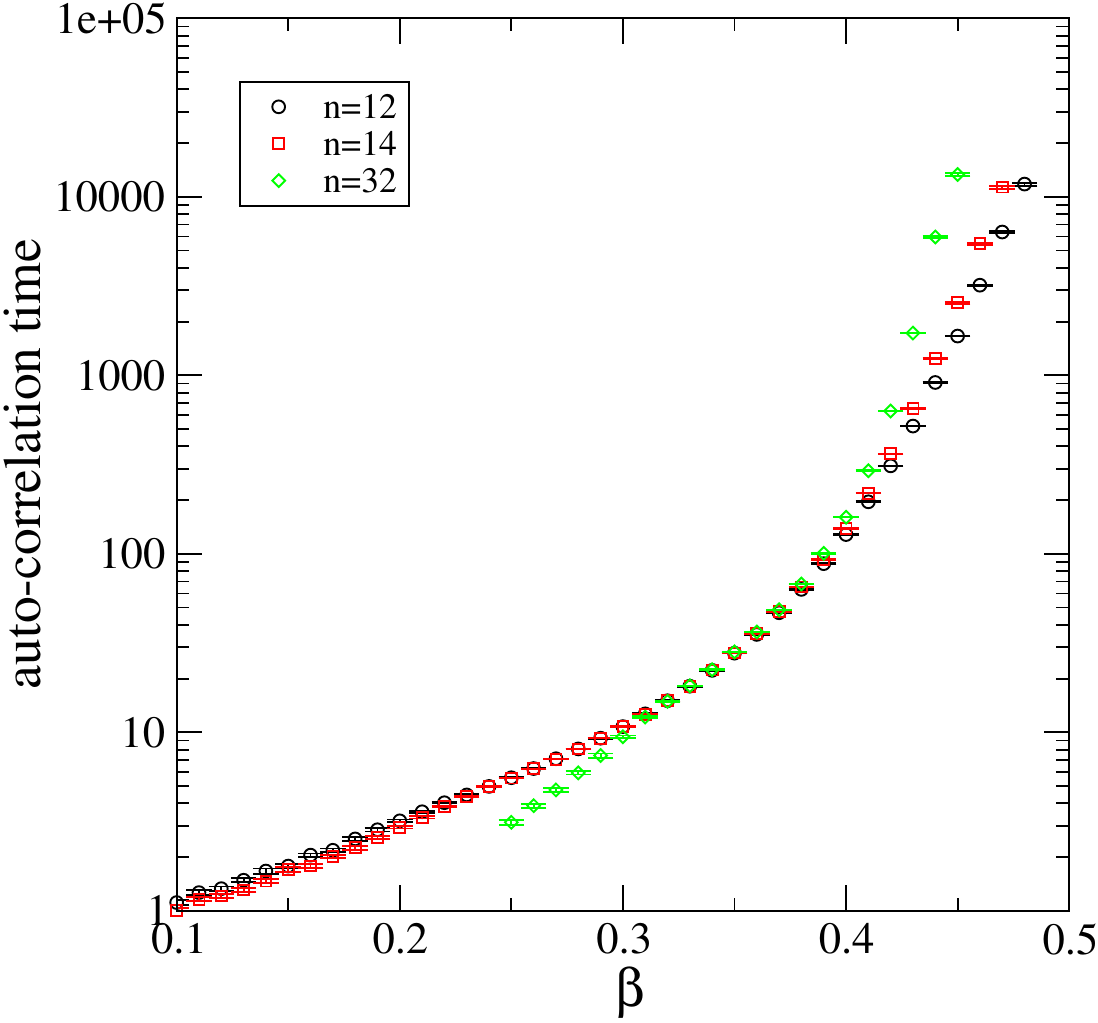}
  \caption{\label{fig:2} Left panel: Solid lines are estimates (see (\ref{eq:23}))
    for the statistical  error $\epsilon $ as a function of the length
    $N$ of the MCMC time series; $12 \times 12$ Ising model. Open
    symbols are the theoretical 
    prediction (\ref{eq:10}). Right panel: extracted autocorrelation
    time as a function of $\beta $ for several lattice sizes;  $n
    \times n$ Ising model. 
  }
\end{figure*}

Partition function and action are invariant under a $Z_2$
transformation of the spins:
\be
s_x \, \longrightarrow  \, (-1) \, s_x \hbo \hbox{for} \; \; \;
\forall x \; .
\label{eq:2}
\en
This means that the
configurations $\{s_x\}$ and $\{ - s_x \}$ have the
same probabilistic weight implying for any {\it finite } lattice size
$V$:
$$
\langle m \rangle  =  \langle s_x \rangle  =  - \langle s_x
\rangle  =  - \langle m \rangle ,   \; \; \Rightarrow \; \;  \langle m \rangle =0
.
$$
It also implies that, during the generation of the MCMC chain, the
sequence
$$
m_1 \to m_2 \to \ldots m_N \; \; \; \hbox{and} \; \; \;
-m_1 \to -m_2 \to \ldots-m_N \
$$
occur with equal probability , meaning the average over chains vanishes as
well, i.e.,
$$
[m] (N) \; = \; 0 \; .
$$

\medskip
The above symmetry enables us to cast each configuration of
the MCMC chain into $Z_2$ classes. To this aim, we define
\be
m_i \; = \; z_i \; \vert m_i \vert \; , \hbo z_i = \pm 1 \; .
\label{eq:4}
\en
Thus, the mapping
$$
\{s\}_i \, \longrightarrow  \, z_i
$$
assigns a $Z_2$ sector (by virtue of the value of $z_i$) to  each
configuration. The symmetry transformation (\ref{eq:2}) maps each
configuration onto a configuration with {\it equal} statistical weight
of the other $Z_2$ sector.

\medskip
The above conclusions are not necessarily true in the {\it infinite
  volume } limit $V \to \infty $. For infinite systems, the $Z_2$
symmetry ca be {\it spontaneously broken}. In fact, the Ising model is
a prototype to explore this phenomenon. For $\beta > \beta _c$, the
statistical system ``freezes'' in one of the $Z_2$ sectors with
$\langle m \rangle \not=0 $. For $\beta < \beta _c$, we still find
$\langle m \rangle =0$ and the symmetry is realised.
The critical value $\beta _c$ can be
calculated analytically~\cite{Onsager}, and one finds:
\be
\beta _c \; = \; \frac{1}{2} \, \ln \left( 1 + \sqrt{2} \right) \;
\approx \; 0.440686 \ldots \; .
\label{eq:5a}
\en
This phenomenon is called {\it spontaneous symmetry breaking} and only
applies to infinite volume systems.

\medskip
Why should we be concerned with this phenomenon since we are only
dealing with cases where $V$ is finite? The answer is that most
importance sampling algorithms (if not all) for large enough $\beta
\gg \beta_c$ and system size $L$, anticipate this phenomenon leading to the
wrong result
$$
[m] (N <N_c) \; \not= \; 0
$$
even at finite size $V$. The theorem of large numbers only
guarantees $[m]=0$ for $N \to \infty $, and on some practical
applications $N_c$ can be unfeasibly large.

\bigskip
Let us study this statement in  the context of an actual numerical
simulation. We generate a chain for the magnetisation
$m_i$ and for the $Z_2$ element $z_i$ as a function of the
Monte-Carlo time $k$ for $\beta = 0.35$ and $L=12$. We observe that
system changes between $Z_2$ sectors during the run, which is expected
since the $Z_2$ symmetry is unbroken at such small values of
$\beta$. However, we realise that regions of positive (negative) $m_i$
cluster for some time. This indicate that we observe a significant
autocorrelation time $\tau $ even at this small $\beta $. In order
to quantify this, we present estimators for the
autocorrelation functions for
$$
m_k, \; \; \;  z_k  \; \; \;\hbox{and} \; \; \; \vert m_k \vert \; .
$$
Note that averages  for $[  m_k ]$ and $[ z_k ]$ vanish but that for $[ \vert
m_k \vert ]$ is non-zero due to the (semi-)positive nature of the
observable.
The simulation is carried out for a $12\times 12$ lattice at $\beta =
0.35$, which is well placed within the symmetric phase with  a moderate
autocorrelation time. The simulation starts with a random spin
configuration (hot-start) and initially discards $1000$
configurations for thermalisation. The
result for the autocorrelation functions is shown in
figure~\ref{fig:1}, right panel. Our findings suggest that the
autocorrelation functions of $m$ and $z$ are proportional (at least
for sufficiently large a MC-time difference), i.e.,
\be
[m_i m_k ] \; \approx \; m_z^2 \; [ z_i z_k ] \; ,
\label{eq:20}
\en
where $m_z^2$ is a parameter, which can be obtained comparing the fits
in figure~\ref{fig:1}, right panel, and which is about $0.149$. This
finding signals that the autocorrelation of the centre sector drives
the overall autocorrelation of the magnetisation.

\medskip
We have systematically studied the error $\epsilon$ (as given by the equation (\ref{eq:7})) for a
$L=12$ lattice size and the three $\beta $ values $0.3$, $0.35$ and
$0.44$.
We fitted the theoretical expression for $\epsilon $ from
(\ref{eq:10}) (the square root of (\ref{eq:10}) to be precise) to the
numerical data. This yields an estimate for $m_0^2$ and the desirable
autocorrelation time $\tau $. Our findings are summarised in
figure~\ref{fig:2}, left panel. For beta $0.3 $ and $0.35$ the
observed autocorrelation time is small
enough so that we can observe the characteristic $1/\sqrt{N}$ behaviour
at large $N$. Note, however, that close to $\beta \approx \beta _c$,
we observe a large autocorrelation time, which does not allow for the
characteristic falloff for the range of $N$ explored. Note, however,
that we still can get an estimate for $\tau $ by virtue of
(\ref{eq:10}), which does {\it not} assume $N \gg \tau $.

\medskip
The same Figure~\ref{fig:2}, right panel, shows the autocorrelation
time as a function of $\beta $ for the three lattice size $12$, $14$
and $32$. We observe that the autocorrelation time increases {\it
  exponentially } in all cases. Note, however, that the slope of the
increase changes around $\beta \approx \beta _c$ and is ``steeper''
for $\beta > \beta _c$, which corresponds to the symmetry broken phase
in the infinite volume limit. 

\medskip
Equation (\ref{eq:20}) suggests that tunneling between $Z_2$ sector is
suppressed and that this suppression is at the heart of the practical
ergodicity issue. For each step in of the MCMC chain, we can assign a
probability $p$ that the configuration changes the $Z_2$ sector during
this step. We then can calculate the autocorrelation $[z_i z_k]$
analytically.

\medskip
In a time series of $k+1$ samples $z_i$, $i=1\ldots k+1$ assume that
$\ell$ transitions occur at $k$ possible locations (links between $i$
and $i+1$). The probability for this event is given by
$$
\left( \begin{array}{c} k \cr \ell \end{array} \right) \; p^\ell \;
(1-p)^{k-\ell} \; .
$$
The contribution of this event to the autocorrelation function
$\langle z_1 z_{k+1} \rangle $ is $(-1)^\ell$. Hence, we find
\bea
\langle z_1 z_{k+1} \rangle &=&\sum_{\ell}^{k}
\left( \begin{array}{c} k\cr \ell \end{array} \right) \; p^\ell \;
(1-p)^{k-\ell} \; (-1)^\ell
\nonumber \\
&=& (1-2p)^{k} \; .
\label{eq:21}
\ena
Using the latter result in (\ref{eq:20}) and exploiting the connection
to the autocorrelation time in (\ref{eq:9}), we find the connection
between autocorrelation time $\tau $ and sector tunneling probability
$p$:
\be
p \; = \; \frac{1}{2} \; \left( 1 - \e ^{-1/\tau } \right) \; \approx
\; \frac{1}{2 \tau } \; .
\label{eq:22}
\en
The latter approximation holds for $\tau \gg 1$. For the example of
the previous subsection, i.e., the heat-bath algorithm, a $12\times
12$ lattice and $\beta = 0.35$, we found $\tau \approx 28$ leaving us
with a tunneling probability of just $p \approx 1.8 \, \%$.

\subsection{Computational resources and precision}

\begin{figure}
  \includegraphics[height=8cm]{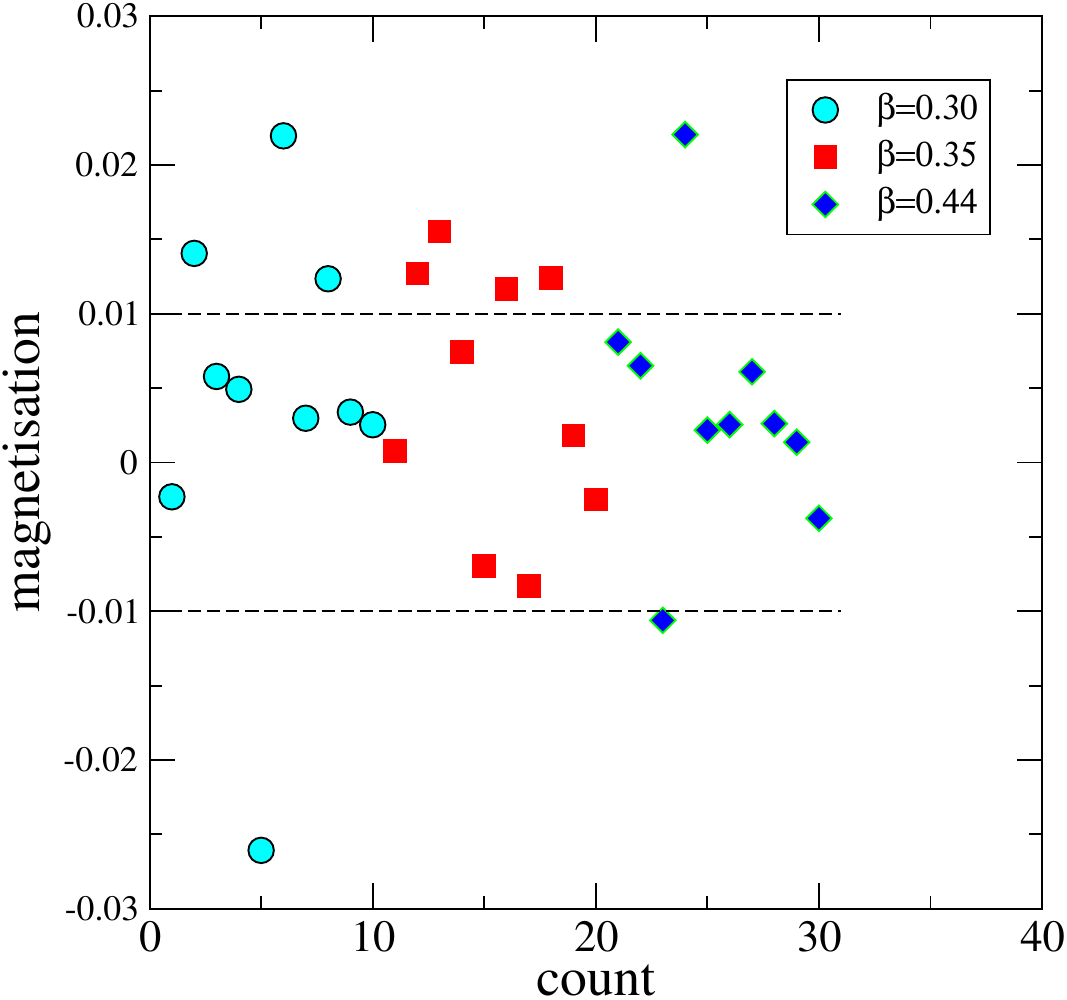}
  \caption{\label{fig:3} Average magnetisation from a MCMC time series
    of length $N$ for three $\beta $ (see~(\ref{eq:24}) for the $\beta
    $-$N$ pairs); $12 \times 12$ Ising model.
  }
\end{figure}
 The strategy of comparing the performance of two different algorithms
is as follows: we will agree at certain level of error $\epsilon^2 $ and
then ask the question how many ``lattice sweeps'' $N$ do we need to
achieve this.

\medskip
For the heatbath algorithm, we already worked out a connection between
$\epsilon ^2$ and $N$ (see 
(\ref{eq:10})), and it depends on only two parameters, i.e., $m_0$ and
$\tau $. It is time to put this equation to the test. We have
generated a time series of $6,000,000$ magnetisations $m_k$, which we divide
into subsequences of length $N$. For each subsequence, we calculate
the average magnetisation
$$
m^{(\alpha)} \; = \; \frac{1}{N} \sum _{k=1}^N m^{(\alpha)} _k ,
$$
where $\alpha $ numbers the subsequences from $1$ to $n_\alpha $,
which fit into the series of $6,000,000$ magnetisations. The error for the
magnetisation estimator (\ref{eq:7}) is then estimated by
\be
\epsilon ^2 (N) \; \approx \; \frac{1}{n_\alpha} \sum _{\alpha
  =1}^{n_\alpha} \left[ m^{(\alpha)} \right]^2 \; .
\label{eq:23}
\en
Our numerical findings for $N=10 \ldots 1500$ appear in
figure~\ref{fig:2}, left panel, as solid lines. We show results for
$\beta = 0.3$, $\beta = 0.35$, $\beta = 0.44$. Each curve is fitted by
the theoretical prediction (\ref{eq:10}) with respect to only two fit
parameters: $m_0$ and $\tau $. The agreement is excellent.

\medskip
We can now ask the question: al least how many MCMC configurations do we need to
achieve $\epsilon  < 0.01$. For an answer, we use (\ref{eq:10}) with
the readily obtained fit parameter $m_0$ and $\tau $. The agreement
between theory and numerical data is that good that we can extrapolate
to $N$ values bigger than $1500$. We find that for our lattice size $L = 12$, $N$ has at least to be:
\bea
\beta &=& 0.30: \; N = 10,800   \label{eq:24} \\
\beta &=& 0.35: \; N = 58,300  \nonumber \\
\beta &=& 0.44: \; N = 10,460,000  \; .
\nonumber
\ena
Note that the above $N$ values are vastly outside the fitting range of
$N=10 \ldots 1500$ and the application of (\ref{eq:10}) is an
extrapolation. It is therefore in order to check the predictions
(\ref{eq:24}). To this aim, we have created, for each $\beta $, an
MCMC time series of length $N$ and have calculated the corresponding
average magnetisation. We have repeated this $10$ times. Since
$\langle m\rangle =0$, we expect these $m$ values to be scattered
around zero with an error band $\epsilon = 0.01$ (one standard
deviation). Our result is shown in figure~\ref{fig:3}. We observed the
expected behaviour even for $\beta =0.44$, for which $N=10,460,000$.

\medskip
It appears that fitting $\epsilon $-data with (\ref{eq:10}) is an
economical way to calculate the autocorrelation time. We have done
this for a range of $\beta $ values and show the result in
figure~\ref{fig:2}, right panel. We observe that the autocorrelation
time $\tau $ exponentially increases with $\beta $. In the ``symmetric
phase'' $\beta \ll 0.44$, the slope seems to be independent of the
lattice size $L$. In the ``broken phase'' $\beta > 0.44$, the picture
changes: the slope of the exponential increase depends on the volume
and is significantly bigger than in the symmetric phase. This signals
a breakdown of validity of the heat-bath simulation for reasonable
sized sample sizes $N$.

\subsection{Volume dependence and Critical Slowing Down}
\begin{figure}
  \includegraphics[height=7.5cm]{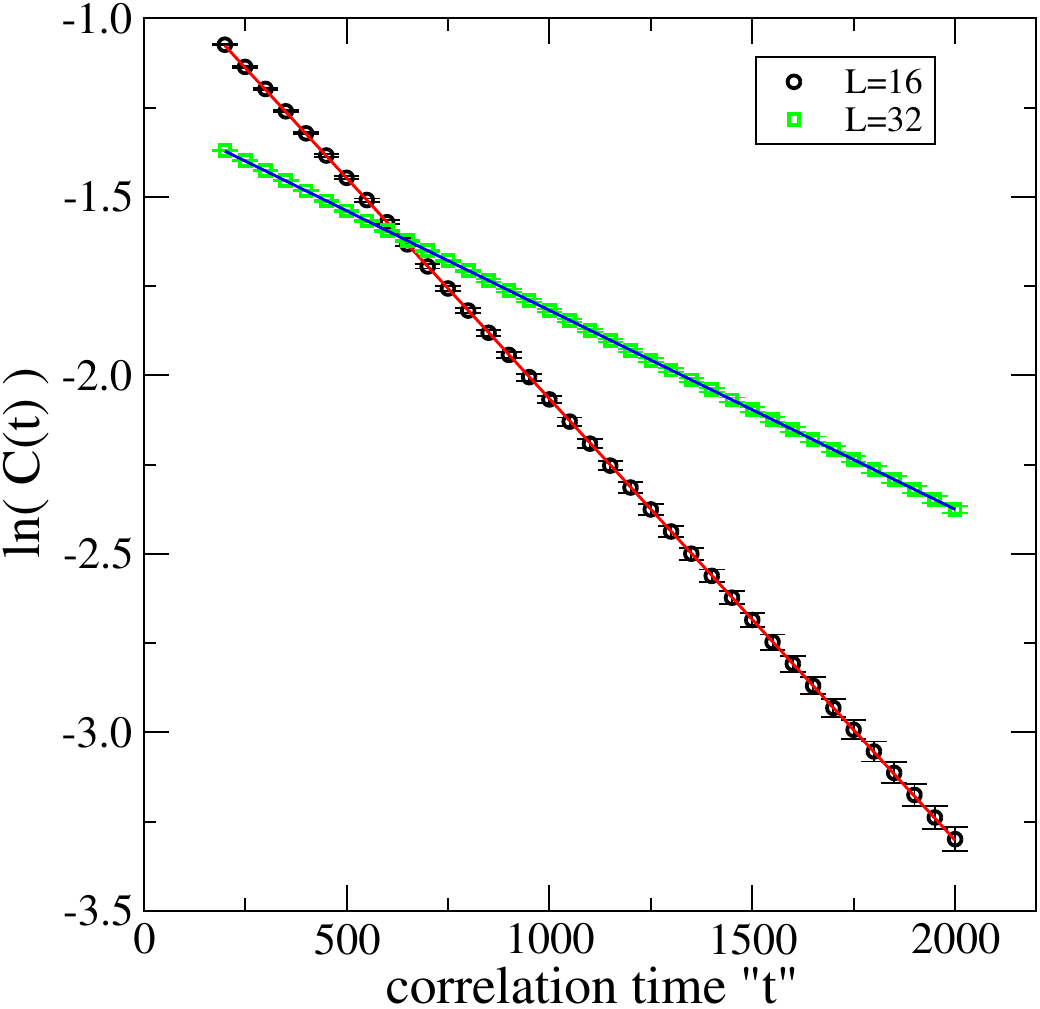} \hspace{0.5cm}
  \includegraphics[height=7.5cm]{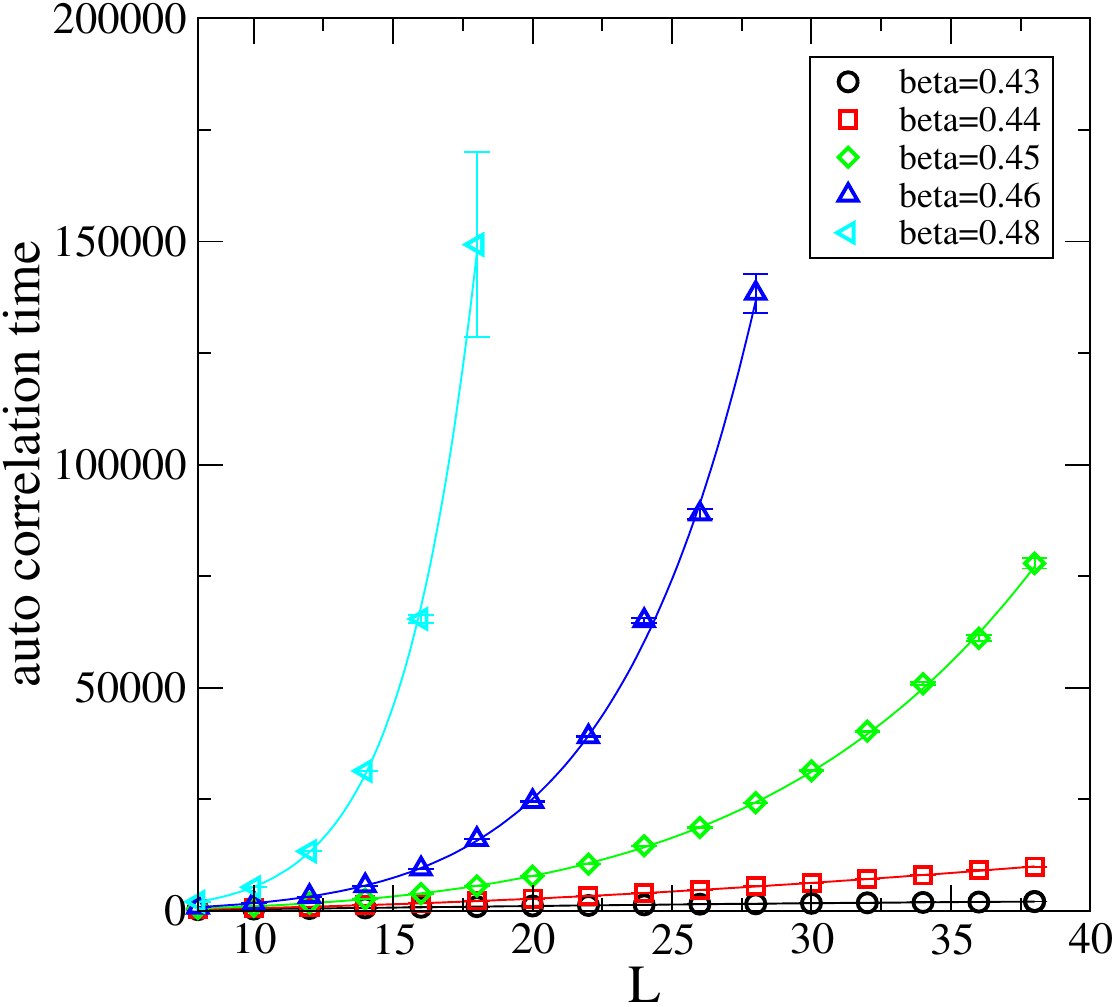}
  \caption{\label{fig:33}  Autocorrelation function as a function of
    Monte-Carlo time $t$   for two lattice sizes at $\beta = 0.43$
    (top). The autocorrelation time for the magnetisation as observable
    as a function of system size $L$ for several values of $\beta $ (bottom). }
\end{figure}
\medskip
Of particular interest is to study the volume dependence of the autocorrelation time at give value of $\beta $. For subcritical values, i.e., $\beta < \beta _c$, we expect a power-law increase with the system size. This is simply because of that we operate with a local update algorithm, for which it is increasingly difficult to disorder a lattice configuration with increasing size. In the broken phase, i.e., $\beta > \beta _c$, the picture is entirely different: the tunneling between centre-sectors is exponentially suppressed and a changing a $Z_2$ sector needs resources with exponentially increase with volume. In this subsection, we will verify this picture with unprecedented numerical evidence.

\medskip
For extracting the autocorrelation time $\tau $ for given size $L$ and $\beta $, we calculate the autocorrelation function as a function of the Monte-Carlo time $t$. We fit the asymptotic tail to a the exponential form:
$$
C(t) \; = \; [m_0 m_t ] \; \propto \; \exp \{ - t/\tau \} \; .
$$
For small $t$, we expect power-law corrections to the above functional form and, for large $t$, the signal might be drowning in the statistical noise of the estimator.
Let $E(t)$ be the estimated error of the function $C(t)$ at time $t$.
For the parameters $L$, $\beta $ explored in this section, we only take data into with
$$
t > 200, \; \hbo t< t_\mathrm{max}, \; ,
$$
where
$$
t_\mathrm{max} : \; \hbox{ largest $t$ with: }  C(t)> 5 E(t)
$$
or $t_\mathrm{max} =2000$ whatever is smaller. This is necessary to keep memory usage under control during the simulation. One of our many results is shown in figure~\ref{fig:33}, top panel. Parameters have been $L=16,32$ and $\beta = 0.43$.
Not all data are shown since the figure would become too crowded.
The numerical data is well fitted by exponential form. Throughout, we monitor the $\chi^2$ of the fit. Errors for the fit parameter and hence the autocorrelation time is obtained by bootstrap. For the fits shown in figure~\ref{fig:33}, we obtained specifically
$$
\tau (L=16) \; = \;  808.5(6) \;, \hbo
\tau (L=32) \; = \; 1794(1) \; .
$$
We have repeated this analysis for $L \in [8,39]$ and $\beta = 0.43, \, 0.44, \, 0.45,\, 0.46, \, 0.48 $. The results for the autocorrelation time $\tau $ is shown in the same figure~\ref{fig:33}, bottom panel. We observe that $\tau $ rapidly grows for $\beta $ values instigating spontaneous symmetry breaking. We observe that the numerical data for $\tau $ are well fitted by the formula
\be
\tau (L) \; = \; b_0 \; L^{b_1} \; \exp \{ b_2 \; L \} \; .
\label{eq:25}
\en
In the absence of the exponential ($b_2=0$), the formula describes a power-law growth of $\tau $ with size $L$ while, for $b_2>0$, the formula suggests an dominating exponential growth. The fits are also shown in the bottom panel of figure~\ref{fig:33}. They well describe the data. In particular, we find:

\begin{table}
  \centering
  \begin{tabular}{crrr}
  & $ \; \; \; \; \; \; \ln(b_0) \; \; \; \; \; \; $ & $\; \; \; \; \; \; b_1 \; \; \; \; \; \; $ & $\; \; \; \; \; \; b_2 \; \; \; \; \; \; $ \cr \hline
  $\beta = 0.43$ &  $1.727(8)$   &  $1.991(4)$   &  $-0.0035(2) $   \cr
  $\beta = 0.44$ &  $1.213(7)$   &  $2.303(3)$   &  $-0.0087(1)$  \cr
  $\beta = 0.45$ &  $ 1.26(3)$    &  $2.26(2)$     &  $0.047(1)$    \cr
  $\beta = 0.46$ &  $1.5(1)$       &  $2.00(7)$     &  $0.130(4)$    \cr
  $\beta = 0.48$ &   $1.5(10)$    &  $1.8(7)$       &  $0.28(6)$   \cr
\end{tabular}
  \caption{Results of the fitting of the lattice size dependence of the autocorrelation time in Monte-Carlo simulations with heatbath updates with a product of power law and exponential functions (\ref{eq:25}).}
  \label{tab:fit_results_heatbath}
\end{table}

\bigskip
We thus find evidence that $b_2$ starts growing to non-zero values around the critical values $\beta \approx \beta _c$ for the phase transition. In the symmetric phase at $\beta =0.43$, we find that the autocorrelation time $\tau $ approximately grows with the volume $L^2$.

%----------------------------------------------------------------------

\section{Reduced critical slowing down with the LLR method}

\subsection{Brief introduction to the LLR approach }

\begin{figure*}
  \includegraphics[height=7.5cm]{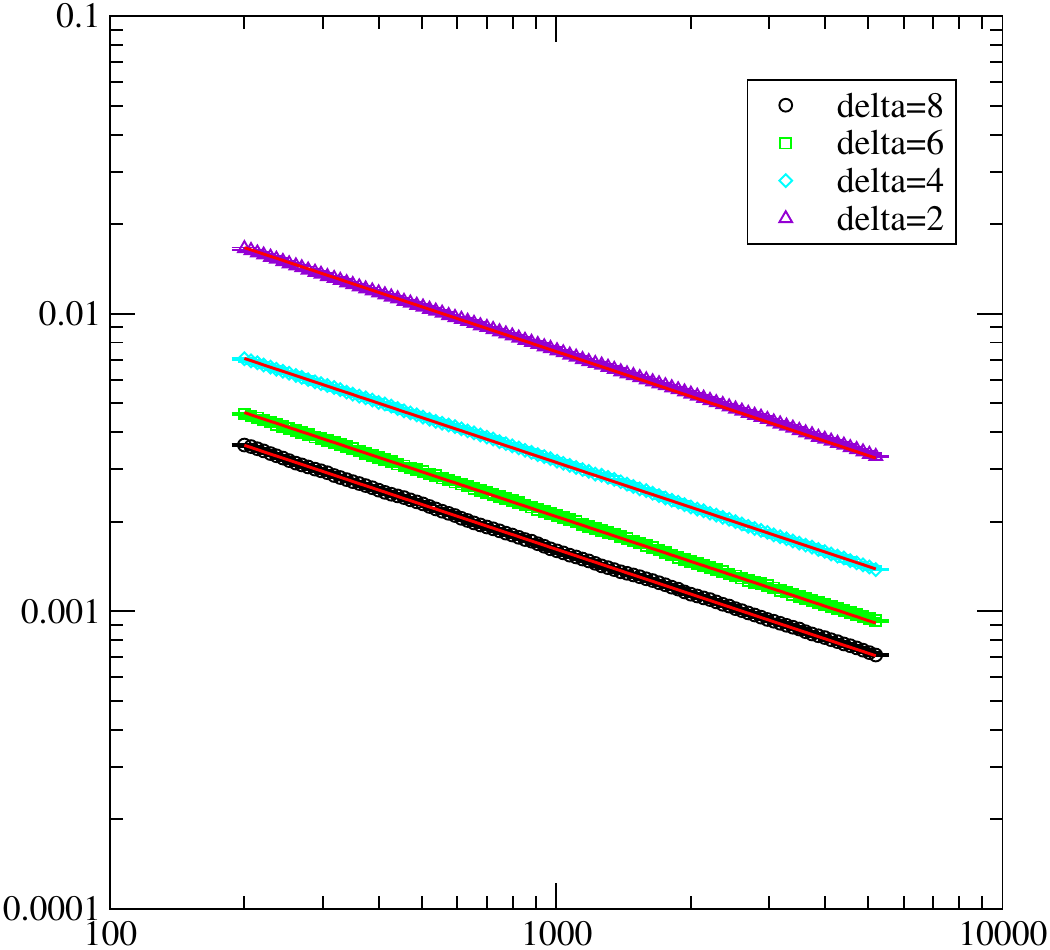} \hspace{0.5cm}
  \includegraphics[height=7.5cm]{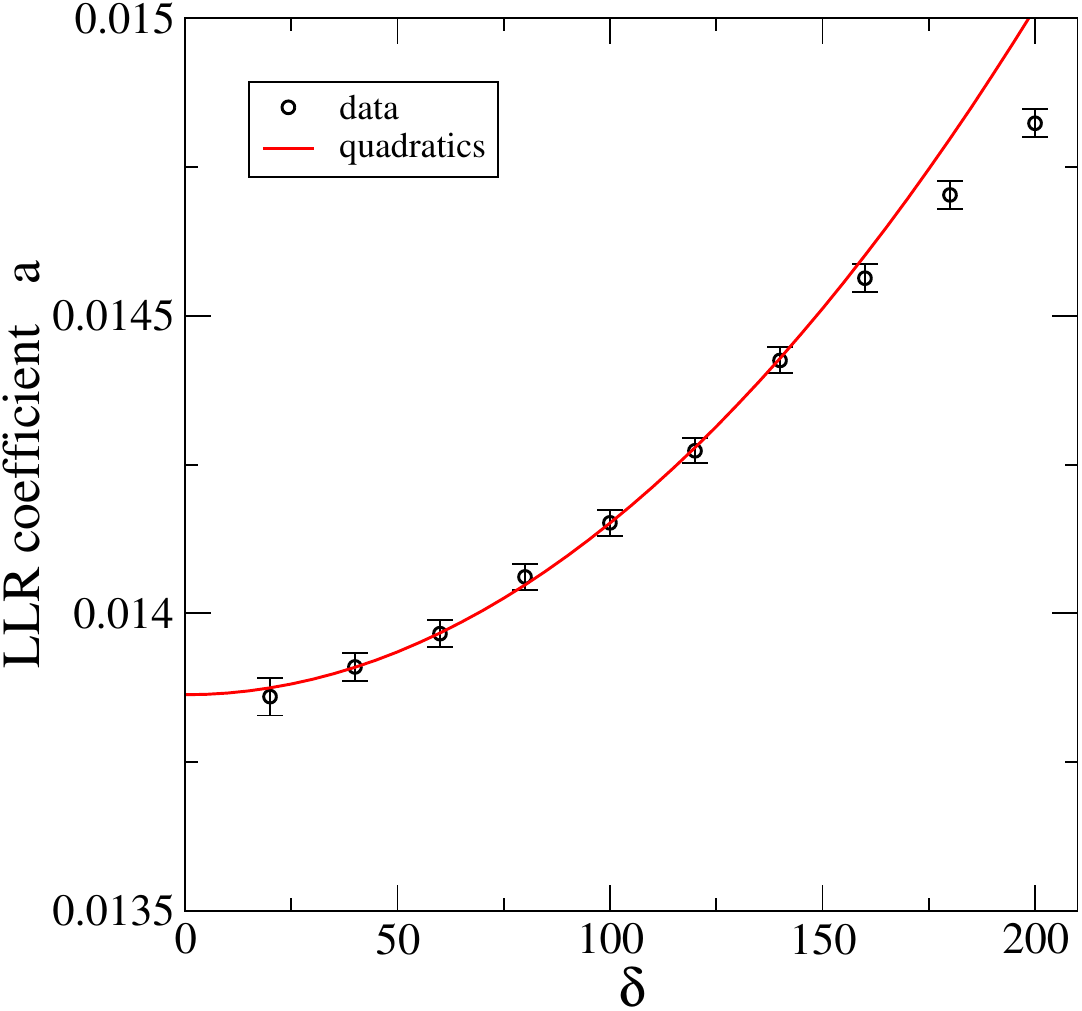}
  \caption{\label{fig:4} Left: The error in the LLR coefficient $a$ as a
    function of the number of Robbins-Monro iterations $n$
    (\ref{eq:37}). The fits correspond to a
    $1/\sqrt{n}$ power law. $12 \times 12$ Ising model, $\beta =
    0.30$. Right: Dependence of the LLR coefficient $a$ on 
    $\delta $ for a $64 \times 64 $ lattice near criticality ($\beta =
    0.44$).
  }
\end{figure*}
 We are aiming to estimate the magnetisation $M$ with reliable errors
over a wide spectrum of $\beta $-values stretching from the
symmetric phase deep into the symmetry broken phase for $\beta \gg
0.44$. We start by defining the density-of-states $\rho (M)$ for the
magnetisation:
\be
\rho (M) \; = \; \frac{1}{Z} \sum _{\{s_x\}} \delta \Bigl( M, \sum _x s_x \Bigr)
\; \exp \{ \beta S\}
\label{eq:30}
\en
with the action $S$ in (\ref{eq:1}). The Kronecker delta is defined
in the usual way:
$$
\delta (i,k) =1 \hbo \hbox{for} \; i=k \; , \hbo 0 \; \hbox{else.}
$$

The magnetisation is then given by
\begin{eqnarray}
\langle m \rangle \; = \; \frac{\sum _M M \; \rho (M) }{ \sum _M \rho
  (M) } \; ,
  \nonumber \\
  M=-V, -V+2, \ldots ,V-2, V.
\label{eq:31}
\end{eqnarray}
With the normalisation
\be
\sum _M \rho (M) \; = \; 1
\label{eq:31b}
\en
because of the definition (\ref{eq:20}) and that of the partition
function $Z$ in (\ref{eq:1}), $\rho (M) $ can be interpreted as the
probability with which magnetisations $M$ contribute to expectation
values such as the one in (\ref{eq:31}). By virtue of the $Z_2$
symmetry transformation (\ref{eq:2}), the density is symmetric, i.e.,
$$
\rho (-M) \; = \; \rho (M) \; ,
$$
leading to $\langle m \rangle =0$ as expected. In our numerical study
we will {\it not } exploit the above symmetry relation but rather will
study the stochastic errors for our estimate for $\langle m \rangle $.

\medskip
At the heart of the LLR approach is the expectation value
\bea
\lb f \rb  (a) &=& \frac{1}{\cal N} \sum _{\{s\}} f(s) \; \e^{ \beta
  S \, + \, a \, m(s) } \; W_\delta \Bigl( m_0, m(s) \Bigr) 
\label{eq:32} \\
m(s) &=& \sum _x s_x \; ,
\nonumber
\ena
where we here use a Heaviside function for the window function:
\be
W_\delta \Bigl( m_0, m(s) \Bigr) \; = \; \left\{ \begin{array}{ll} 1 &
\hbox{for} \; m_0 - \delta \le m(s) \le m_0 + \delta \; . \cr
0 & \hbox{else.} \end{array} \right.
\label{eq:33}
\en
Note that $\lb f \rb (a)$ depends also on the parameters $\delta $ and
$m_0$, and $a$ is also called the LLR coefficient. You can obtain the
density-of-states $\rho (m_0)$ by carrying out the following steps:

\begin{enumerate}
  \item[1.] For a given $\delta $ and $m_0$, solve the stochastic
    equation
    \be
    \lb  m(s) - m _0\rb (a^\ast) \; = \; 0
    \label{eq:34}
    \en
    for $a$ (solution $a^\ast$), which depends smoothly on $m_0$ and $\delta $ for $m_0
    \in \left[-V,V\right]$.
  \item[2.] Use
    \be
    \frac{d}{dm_0} \ln \rho (m_0) \; = \; \lim _{\delta \to 0} a
    (\delta, m_0)
    \label{eq:35}
    \en
    and evaluate (or estimate) $ \rho (m_0) $ up to a multiplicative
    factor by integrating the above equation.
  \item[3.] Determine the multiplicative factor by normalising $\rho $
    (see (\ref{eq:31b})).
\end{enumerate}
The last step might be optional since a normalisation constant drop
out of expectation values such as the one in (\ref{eq:30}).

\medskip
As for the heat-bath MCMC approach, we are interested in the question:
what type of precision can we achieve as a function of the invested
computational resources. We therefore will critically investigate the
parameter dependence of the numerical error.

\medskip
Let us first comment on solving the stochastic equation of the type
(\ref{eq:34}). This task has been extensively studied firstly by
Robbins and Monro~\cite{robbins_monro} and then taken up by number of
authors (see \cite{kushner1997} for a review). If $F(a)$
is a noisy estimator for
\be
f(a) := \lb  m(s) - m _0 \rb (a)
\label{eq:35c}
\en
 Robbins and Monro propose an under-relaxed iterative approach. Starting with
some $a_1$, consider the recursion
\be
a_{n+1} = a_n \; - \; \alpha _n \; F(a_n)
  \label{eq:36}
  \en
with a sequence of positive weights $\alpha _n$, $n=1,2 ,3 \ldots $
satisfying
$$
\sum _{n=1}^\infty \alpha_n \to \infty \; , \hbo \sum _{n=1}^\infty \alpha^2_n
\to \hbox{finite} \; .
$$
The sequence converges with probability one to the solution $a^\ast :=
a_\infty $~\cite{Blum1954}.
A particular sequence was suggested by Robbins
and Monro:
$$
\alpha _n = \frac{\kappa }{ n} \; .
$$
The algorithm reaches {\it asymptotically } the optimal convergence
rate of $1/\sqrt{n}$, but the initial (low $n$) performance crucially
depends on the sequence. Chung~\cite{chung1954} and Fabian~\cite{fabian1968}
showed that optimal convergence is reached with the choice:
$$
\alpha _n = \frac{1 }{f^\prime (a^\ast) \;  n} \; .
$$
This choice, however, hinges on the solution $a^\ast $.
For the specific problem at hand, i.e., (\ref{eq:34}), we can,
however, find a good value $\kappa $. For small enough $\delta $, the
marginal for the magnetisation $m$ in the window
$[m_0-\delta,m_0+\delta]$ is Poisson distributed, i.e., $\propto \exp
\{-a^\ast m \}$. Together with the 're-weighting' factor $\exp \{ a m
\}$ in (\ref{eq:32}), the $m$ distribution becomes flat for values $m$
inside the window. We then find with (\ref{eq:35c}), the definition
(\ref{eq:32}) and the solution (\ref{eq:34}):
\bea
f^\prime (a^\ast ) &=&  \lb  (m(s) - m _0) m(s) \rb (a^\ast)
\nonumber \\
&=& \lb  (m(s) - m _0) ^2\rb (a^\ast) = \frac{1}{2\delta +1} \sum
_{m=-\delta }^\delta m^2
\nonumber \\
&=& \frac{\delta \, (\delta+1) }{3} \; \approx \; \frac{\delta ^2 }{3}
\;.
\nonumber
\ena
The latter hold for $\delta \gg 1$, which would also be the result if
the degrees of freedoms have a continuous domain of support. Note that
by the nature of the task at hand (\ref{eq:34},\ref{eq:32}), $f^\prime
(a^\ast )$ does not depend on the solution $a^\ast $. We arrive at the
iteration that we will study in the remainder of the paper:
\be
a_{n+1} = a_n \; - \; \frac{3}{\delta ^2 \; n } \; F(a_n)  \; .
  \label{eq:37}
\en
We put the above iteration to the test for a $V=12\times 12 $
lattice, $\beta = 0.3$, $m_0= INT(0.8 V)$  and several $\delta $ values. The estimator
$F(a)$ is obtained by $20$ successive lattice sweeps. Our findings for
the error $\epsilon _a$ in the LLR coefficient $a$ as a function of the Robbins Monro
iteration time $n$ is shown in figure~\ref{fig:4}. We performed
$1,000$ independent Robbins Monro runs to estimate the error for
$\epsilon _a$. We find optimal convergence behaviour already for
$n>200$. The error for small $\delta $ are smaller than those for
large $\delta $. This is expected since for larger $\delta $ the
window function is wider and hence includes more spin in the
averaging.

\subsection{Precision versus resource }

\begin{figure*}
  \includegraphics[height=7.5cm]{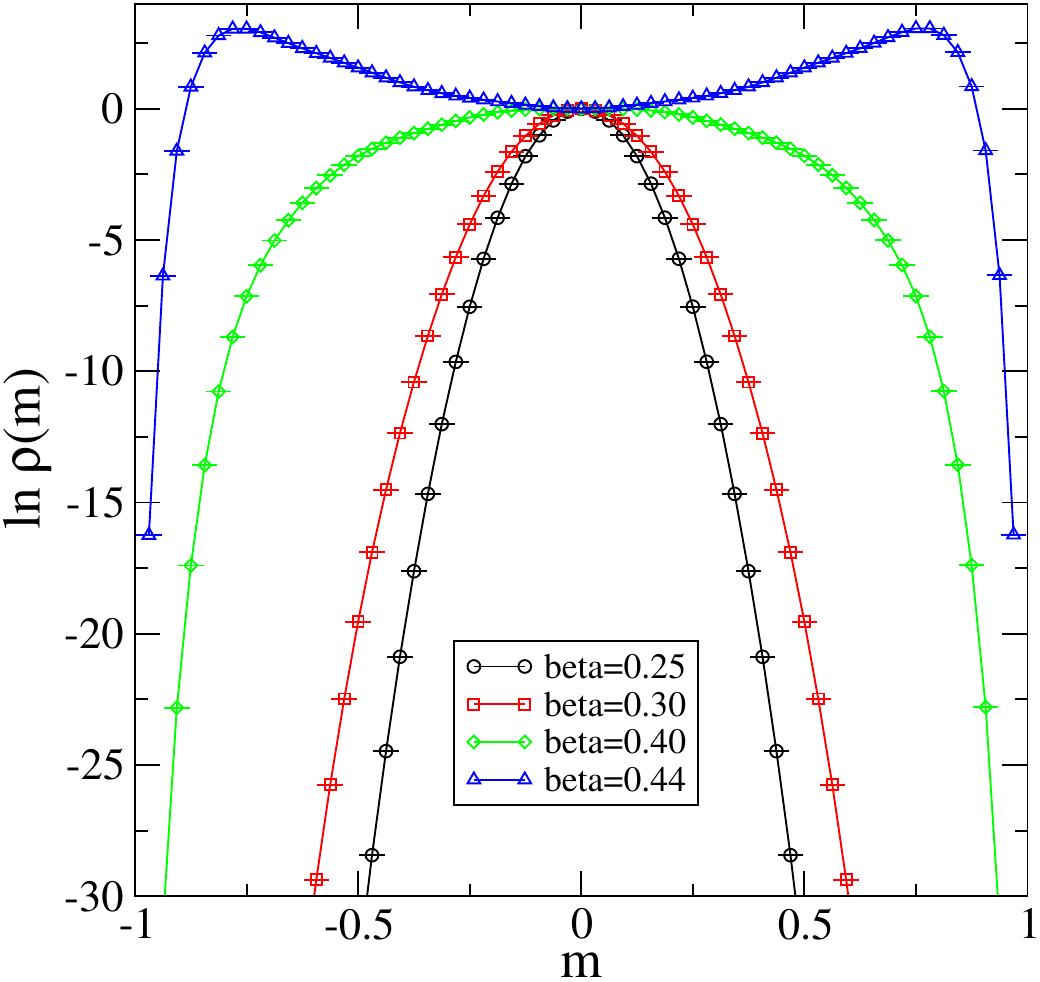} \hspace{0.5cm}
  \includegraphics[height=7.5cm]{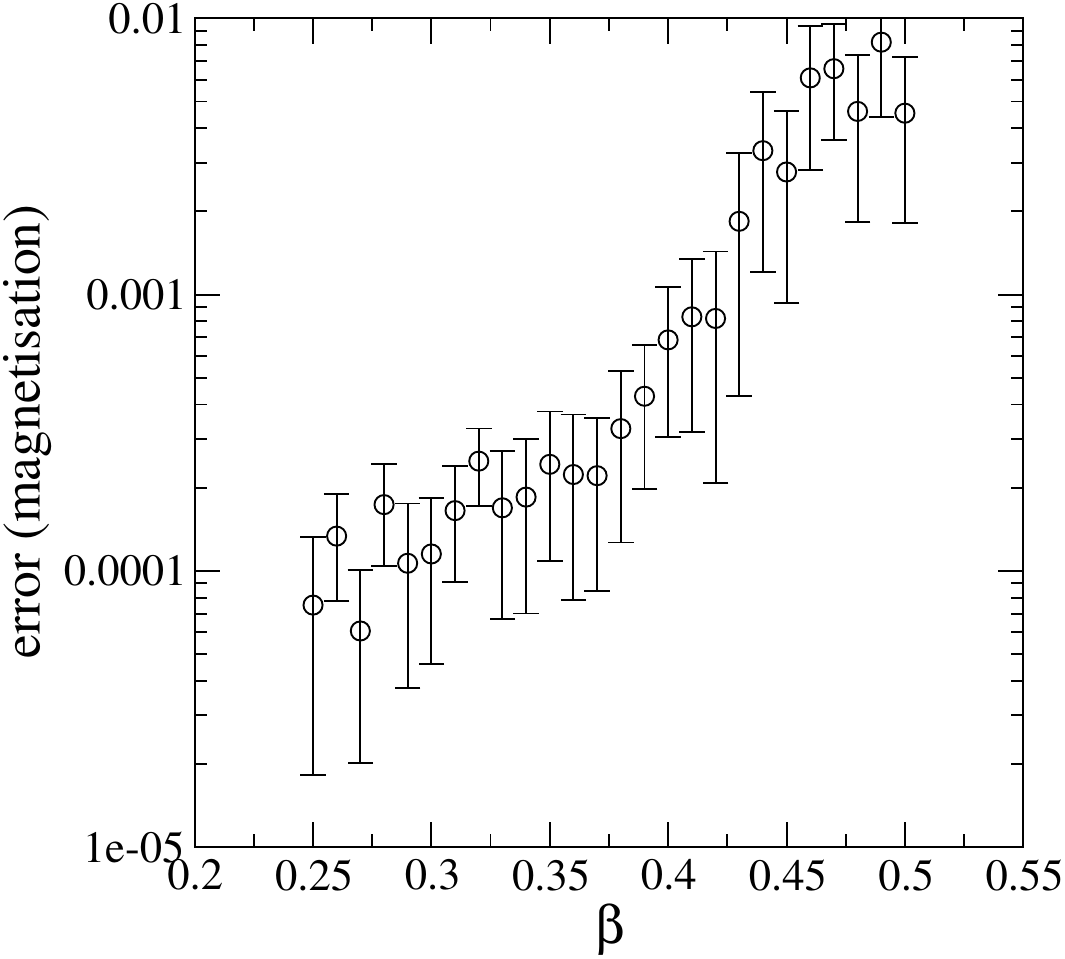}
  \caption{\label{fig:5} Left: Log of the density of states $\rho (m)$ as a
    function  of the (intensive) magnetisation $m$ for four $\beta $
    values; $32 \times 32$ Ising model.  Right: Error of the
    magnetisation (\ref{eq:40}) as a 
    function of $\beta $.
  }
\end{figure*}
The following study is done for the 2D Ising model on a $32 \times 32$
lattice. The objective is to find the amount of 'lattice sweeps' is
needed to calculate the magnetisation $\langle m \rangle $ to a given
accuracy. In the last section, we saw that the heat bath algorithm
needs a rapidly increasing amount of resource if $\beta $ approaches
the regime of a spontaneously broken symmetry.

\medskip
Our simulations parameters are ``ball park'' figures and are {\it not}
fine tuned.

\begin{enumerate}
\item We use a step function as window function $m \in [m_0-\delta,
  m_0 +\delta ]$ with $\delta =8,16,24,32 $.
\item We perform $10,000$ Robbins Monro iterations for each $m_0$ and
  for each $\delta $ leaving us with an estimate for the LLR parameter
  $a(\delta )$. We perform a quadratic fit for extrapolating to $\delta
  \to 0$ and set: $a = a(0)$.
 \item Each double expectation value is estimated with $20$ lattice
   sweeps.
 \item We generate LLR parameters $a$ for $63$ values
   of $m_0$, i.e., $(m_0)_k = - 32^2\, + \, k \, \times \, 32 $, $k=1 \ldots
   63$.
\item For each $m_0$, we generate $80$ potential LLR parameters $a_i$
  for the subsequent statistical analysis.
\end{enumerate}
 We will measure resource in units of 'lattice sweeps' ($ls$), i.e., one
 resource unit corresponds to $V$ spin updates. This choice allows to
 measure resource independent of hardware employed for the
 calculations. All algorithms studied here - heat bath update,
 cluster algorithms, LLR method - uses 'lattice sweeps' at low level
 of the calculation. Although Ising spin updates are low cost, the
 'lattice sweep' might be the most expensive computational element for
 other systems such as gauge theories with fermions (QCD) where a
 lattice sweep could be defined by a Hybrid Monte-Carlo trajectory.

 \medskip
 To generate the above data set for the LLR coefficients (steps 1-4), the resources
 needed are
 \be
 4 \, \times \, 20 \, \times \, 10,000 \times 63 \, ls \; = \; 5.04 \cdot 10^7 \,
 ls \; .
\label{eq:38}
 \en
 From this data set, we can already estimate expectations values of
 functions of the magnetisation, and the objective here is to
 estimate the precision with which we can calculate $\langle m \rangle $
 (which equals zero for a simulation with infinite resources). To
 this aim, we will repeat the calculation $80$ times. This, the
 analysis uses the resources of $5.04 \cdot 10^7 \times 80 \, ls = 4.032 \cdot 10^9 \,
 ls$, which must not be confused with resource (\ref{eq:38}) needed to
 produce one sample result.

 \medskip
The density of states $\rho $ for the magnetisation $m$ is obtained by
integration of the LLR- coefficient:
\be
\rho (m) \; = \; \exp \left\{ \int _0^m a(m^\prime ) \; dm ^\prime
\right\} \; .
\label{eq:39}
\en
The normalisation is arbitrarily chosen to be $\rho
(0)=1$. Expectations values are then obtained by a second integration,
e.g.,
\be
\langle m \rangle \; = \; \int m \; \rho (m) \; dm \; / \, \int  \rho (m) \;
dm \; .
\label{eq:40}
\en
Early studies~\cite{Langfeld:2012ah, Langfeld:2014nta, Langfeld:2015fua}
used a trapezium rule and summation, which leads to an
accumulation of error for increasing $m$. Representing the function
$a(m)$ by high degree polynomial and performing the integrations (semi-)~analytically has proven very successful~\cite{Langfeld:2016kty,
 Garron:2016noc, Garron:2017fta,phi_llr2020}. One can prove that the
density of states for Ising model is an even function in $m$ by virtue
of its $Z_2$ symmetry. Correspondingly, the LLR coefficient
$a(m)$ is an odd function. A numerical approach exploiting this
observation would approximate $a(m \ge 0)$ by polynomial of odd
powers of $m$. This would lead to the exact result $\langle m \rangle =0$.

\medskip
The prime objectives here is to avoid any assumptions on symmetry and
to observe to what extent the exact result $\langle m \rangle =0$ is obtained. For this purpose, we
approximate $a(m)$ over the full domain by polynomial containing even
and odd powers of $m$. We find that a polynomial of degree $16$
represents the numerical data for $a$ very well.

\medskip
The result for $\rho (m)$ (on a logarithmic scale) is shown in
figure~\ref{fig:5}. Error bars are obtained by the bootstrap method:
\begin{enumerate}
 \item For each $m_0$, calculate a set of $n_B$ LLR coefficients from
   independent runs. We have chosen here $n_B=60$.
\item For each of the (discrete) $m_0$ choose an LLR coefficient out
  of the $n_B$ possibilities.
\item Fit a polynomial of degree $16$ to the data.
\item Perform the integration (\ref{eq:39}) analytically and obtain
  one sample for $\rho (m)$.
\item[5a.] Repeat this procedure many times and calculate the average for
  $\rho (m)$ and the standard deviation (error bar).
\end{enumerate}
Step 5a gives rise to the graphs in figure~\ref{fig:5}, left panel. We
find that for $\beta = 0.25, \, 0.30, \, 0.40$ the density-of states
is maximal at $m=0$ making $m=0$ the most likely magnetisation. We
also observe that, for a finite $L=32$ lattice, the curve for $\beta
=0.44$ develops a double peak structure, which is characteristic for
the spontaneous breakdown of symmetry. We expect that for increasing
lattice size, the $\beta $ for which the double peak structure occurs
will approach $\beta _c$ in (\ref{eq:5a}).

\medskip
We are here not primarily interested in the density of states $\rho $
but the expectation value of the magnetisation
$$
m \; = \; M/V \; = \; \frac{1}{V} \sum _x z_x \; .
$$
In this case, we replace step 5a by:
\begin{enumerate}
\item[5b.] For the sample $\rho (m)$, calculate the two integrals in
  (\ref{eq:39}) analytical and, thus, obtain a sample value for $\langle
  m \rangle $. Repeat this procedure many times and calculate the average for
  $\langle m \rangle $ and the standard deviation (error bar).
\end{enumerate}

Figure~\ref{fig:5}, left panel, shows the (log of the) density of
states as a function of the intrinsic magnetisation $m = M/L^2$. For
the finite volume $L=32$, we see that the most likely
magnetisations $m$ are at $m \not=0$ for $\beta=0.44$. This is a
precursor of spontaneous symmetry breaking. Increasing the volume, it is
expected that this bifurcation moves up in $\beta $ to approach $\beta
_c$ (\ref{eq:5a}) in the infinite volume limit.

\medskip
Having calculated the density of states, we estimated the
magnetisation $m$ using (\ref{eq:40}). The precision with which the exact result $\langle m \rangle =0$ is recovered depends on the quality of the symmetry $\rho (m) = \rho (-m)$. Our result for the {\it error} of $m$ is shown in figure~\ref{fig:5}, right panel, as a function of $\beta$, where we
have kept fixed the number of Robbins Monro iterations and the
bootstrap copies. We find a moderate increase with increasing $\beta$, which can be explained by the larger variation of $\rho(m)$ with $m$
due its peak structure, which makes it harder to control the numerical
precision of the integration over $m$ in the integrals of (\ref{eq:32}).

\subsection{Autocorrelations and density-of-states }

\begin{figure}
  \includegraphics[height=8cm]{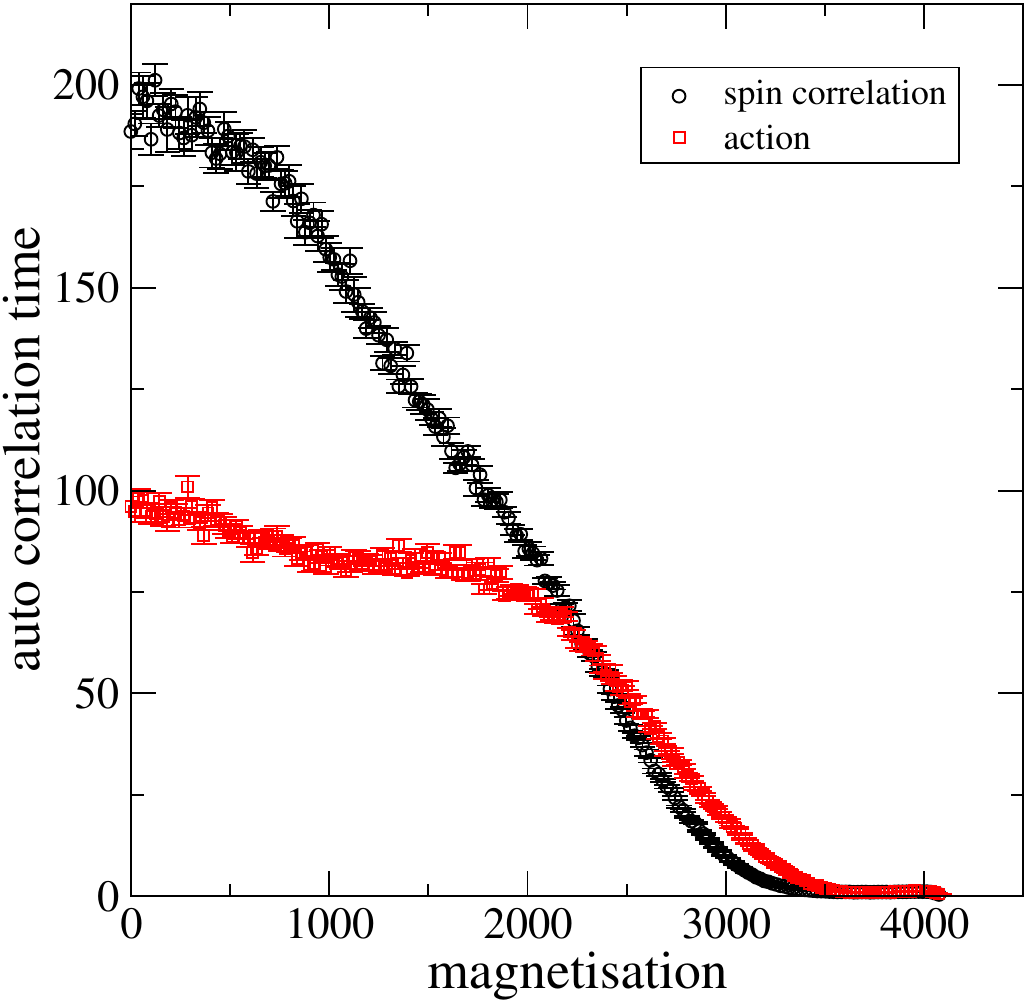}
  \caption{ Autocorrelation time for the LLR double expectation value
    for the action and the spin-spin correlation as a function of
    $m_0$. $64 \times 64$ Ising model, $\beta =0.44$, $\delta =40$. 
  }  \label{fig:6}
\end{figure}
\
The so-called double expectation values such as in (\ref{eq:23}) are
at the heart of the LLR approach since they ultimately give rise to
$a$ and hence the density of states (see (\ref{eq:34}). These
expectation values can be viewed as ordinary Monte-Carlo expectation
values, and, as such, they are susceptible to autocorrelations of the
Markov chain. 

\medskip
We already established that there is a close link between spontaneous
symmetry breaking and the exploding autocorrelation time for local
update algorithms operating close to criticality. We expect that the
double expectation values are much less affected by this phenomenon
simply because they are not operating a close to criticality ``most of
the time''. 

\medskip
We first note that the double expectation values depend on a number of
parameters, which are not present in a standard heat bath
simulation. There is the LLR parameter $a$ which adds a term $a \sum
_x s_x$ to the action. For $a \not=0$ this parameter acts like a
magnetic field, which breaks the $Z_2$ symmetry $s_x \to -
s_x$. Secondly, the window function $W(m_0, m(s) ) $ 
(\ref{eq:33}) is 
part of the probabilistic measure. It restricts spin configurations to
values of the magnetisation $m(s)$ close $m_0$. This means that this
factor also breaks the $Z_2$ symmetry as long as $m_0 \not=0$. Note,
however, that for $m_0 =0$, the solution of the stochastic equation is
$a=0$ precisely because of the $Z_2$ symmetry. We thus expect that the
calculation of $\rho (m \approx 0)$ might be affected by long
autocorrelations. 
Note that for most of the observables in the broken phase, $\rho(m
\approx 0)$ might be an entirely suppressed domain of integration for
the integrals in e.g.~(\ref{eq:40}). In this case, these
autocorrelations have little impact on the precision of the
calculation. 

\medskip
In a first step, we studied the autocorrelation time for the action and the spin-spin correlation function for different values of $m_0$, the centre of the window function:
$$
\hbox{action:  } \sum _{\langle xy\rangle } s_x s_y \; , \hbo \hbox{  spin-spin:  }
s_x \, s_{x+L/2} \; .
$$
Our findings are summarised in figure~\ref{fig:6}, left panel. Indeed, we observe that those autocorrelations are highest close to $m_0=0$ where the system can have critical behaviour.

\medskip
\begin{figure}
  \includegraphics[height=8cm]{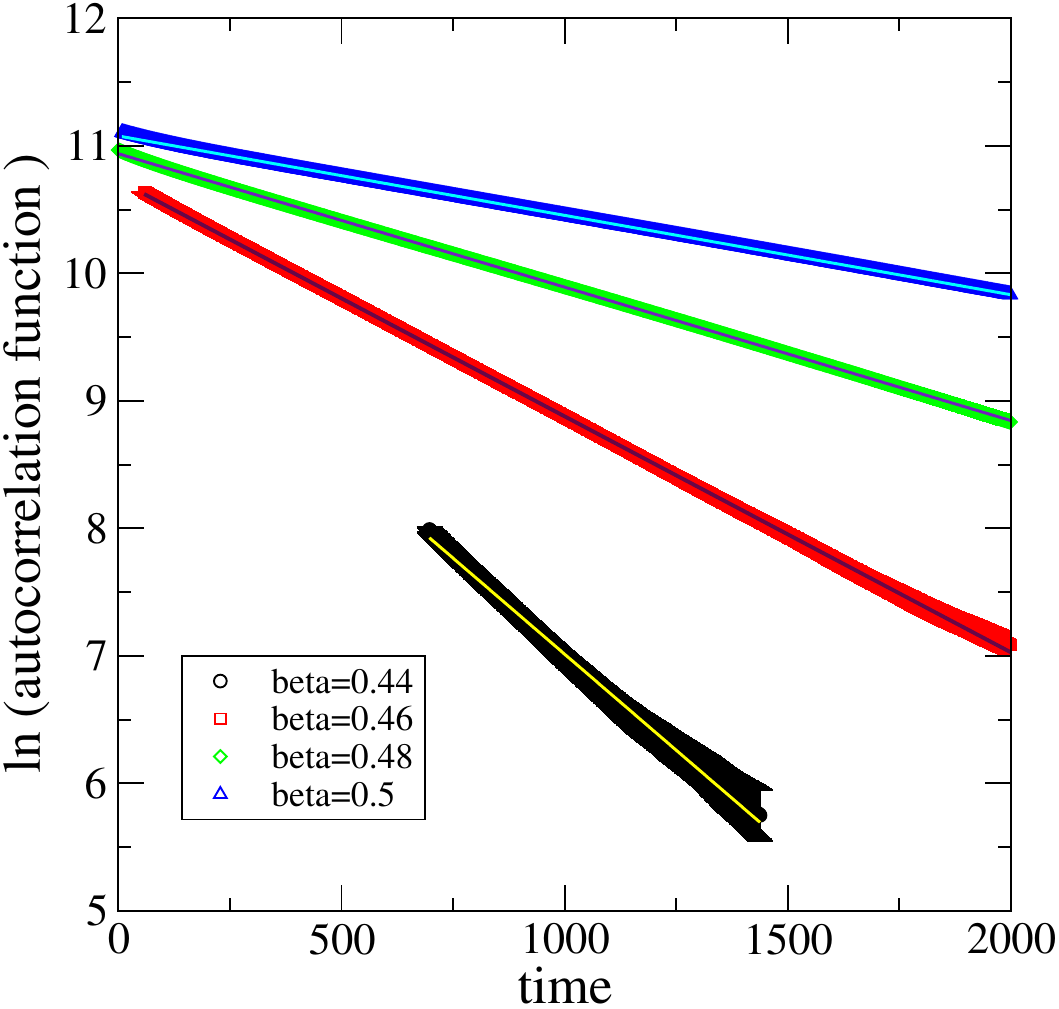}
  \caption{Autocorrelation function for the observable $M_1$
    (\ref{eq:51}) as a function of the autocorrelation time for four
    values of $\beta $. $32 \times 32$ Ising model, $m_0=0$. 
  }  \label{fig:7}
\end{figure}
\begin{figure}
  \includegraphics[height=8cm]{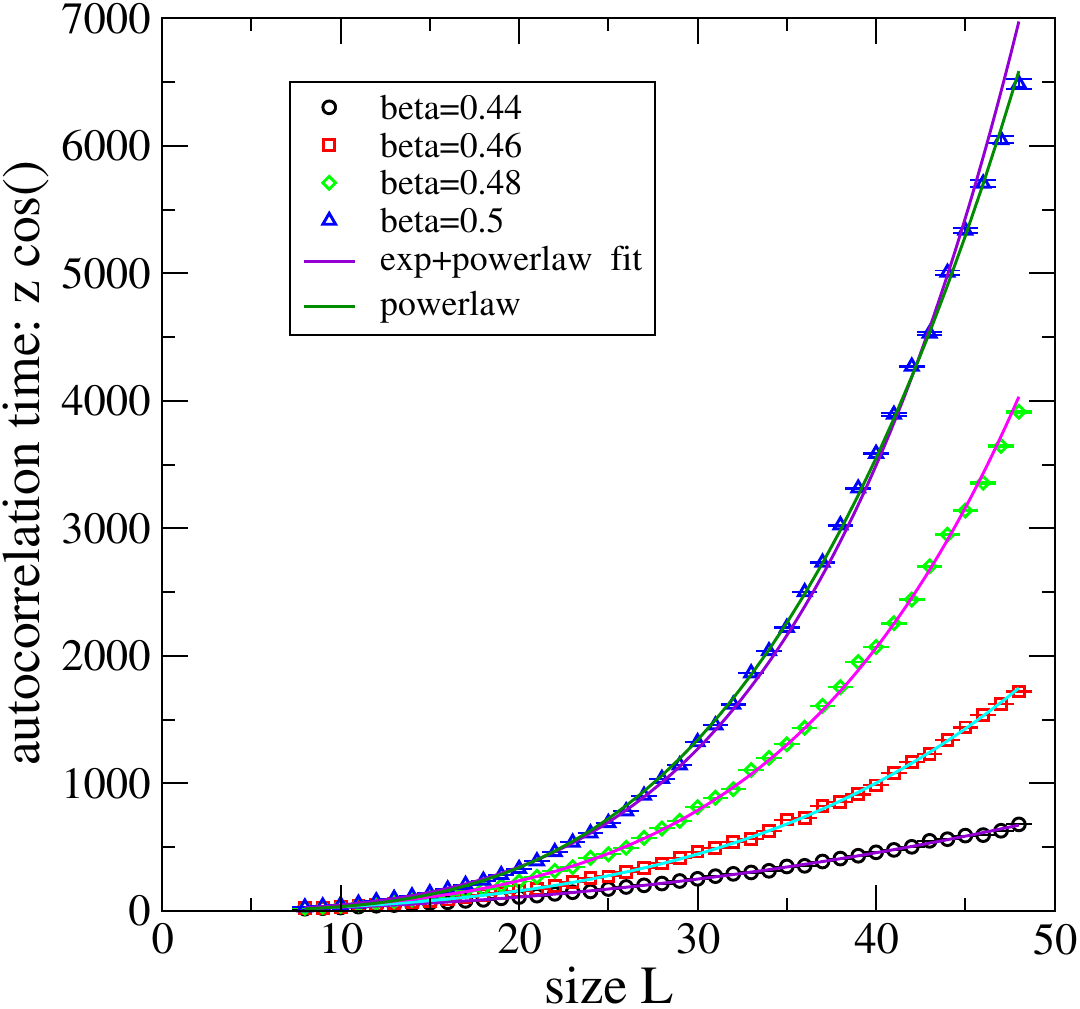}
  \caption{ Autocorrelation time for the observable $M_1$ (\ref{eq:51}) as a function of the system size $L$ for four values of $\beta $ and for the worst case scenario $m_0=0$.
  }  \label{fig:8}
\end{figure}

Since the magnetisation is constrained to a region around $m_0$ in the LLR simulation, autocorrelations of the magnetisation are indeed very small. In search of an observable susceptible to longest autocorrelations, we introduce the Fourier transform of the magnetisation:
\be
 \bar{M}(p_x, p_y) \; = \; \sum_{x} s_{x, y} \;\cos \left( \frac{ 2\pi }{L} \, \lr{x \, p_x + y \, p_y } \right) \; ,
\label{eq:50}
\en
For $p_x = 0, p_y = 0$, this quantity becomes the magnetisation, i.e., $M = \bar{M}(0)$. Another ``infrared'' observable, similarly prone to autocorrelations but unconstrained by the LLR approach, is $\bar{M}$ for the lowest momenta with either $p_x = 1$, $p_y = 0$ or $p_x = 0$, $p_y = 1$. The choice of these observables is motivated by the common observation that low-momentum modes typically have the slowest relaxation/decorrelation rate in local, translationally invariant quantum field theories. We thus study the autocorrelation time for the observable
\be
M_1 \equiv \bar{M}(1, 0) \; = \;  \sum_{x, y} s_{x, y} \;\cos \left( \frac{ 2\pi }{L} \, x \right) \; .
\label{eq:51}
\en
To this end, we firstly estimate the autocorrelation function $C(t)$ of $M_1$ and extract the autocorrelation time by analysing the exponential decrease at large values of $t$. If $t$ is too large, statistical noise drowns the signal. If $\sigma (t)$ is the standard deviation of the estimator for $C(t)$, we only use data with
$$
C(t) \; > \; 5 \, \sigma (t) \; .
$$
At small values of $t$, $C(t)$ is not well represented by an exponential function, which only hold asymptotically. We proceed as follows: starting at $t=t_0=0$, we fit an exponential function to the data and obtain the $\chi^2 /\mathrm{dof}$. We then systematically increase $t_0$ until $\chi^2 /\mathrm{dof}$ falls below $0.8$ for the first time. We thus extract the autocorrelation time $\tau $ from the fit:
$$
a_0 \; \exp \{ - t / \tau \} \; .
$$
Figure~\ref{fig:7} shows the correlations function $C(t)$ for a $32^2$
lattice and for four values of $\beta $ within the dynamically
generated domain of support. Repeating this procedure for lattice
sizes between $L=8 $ and $48$, we find the result shown in
figure~\ref{fig:8}. We indeed observe that the autocorrelation times
for $M_1$ increase with increasing lattice size $L$, but not nearly to
the extent as we have seen those for the heatbath simulation and the
magnetisation $M$. 

The central question is whether or not these autocorrelations times
increase {\it exponentially} with $L$. In search of an answer, we have
employed the same fit (\ref{eq:25}) of the data as in the case of the
heatbath result. Of particular interest is the coefficient $b_2$,
which indicates super critical slowing down for $b_2>0$. our findings
are summarised in the table below: 

\begin{table}
  \centering
  \begin{tabular}{crrr}
  & $ \; \; \; \; \; \; \ln(b_0) \; \; \; \; \; \; $ & $\; \; \; \; \; \; b_1 \; \; \; \; \; \; $ & $\; \; \; \; \; \; b_2 \; \; \; \; \; \; $ \cr \hline
  $\beta = 0.44$ &  $-1.28(1)$   &  $1.965(4)$   &  $0.0038(2)$  \cr
  $\beta = 0.46$ &  $ -1.26(3)$    &  $1.942(3)$     &  $0.025(1)$    \cr
  $\beta = 0.48$ &  $-1.495(6)$       &  $2.080(3)$     &  $0.036(1)$    \cr
  $\beta = 0.50$ &   $-2.194(7)$    &  $2.484(4)$       &  $0.029(2)$   \cr
\end{tabular}
  \caption{Results of the fitting of the lattice size dependence of the autocorrelation time in LLR simulations with a product of power law and exponential functions (\ref{eq:25}).}
  \label{tab:fit_results_LLR}
\end{table}
\begin{figure}
  \includegraphics[height=8cm]{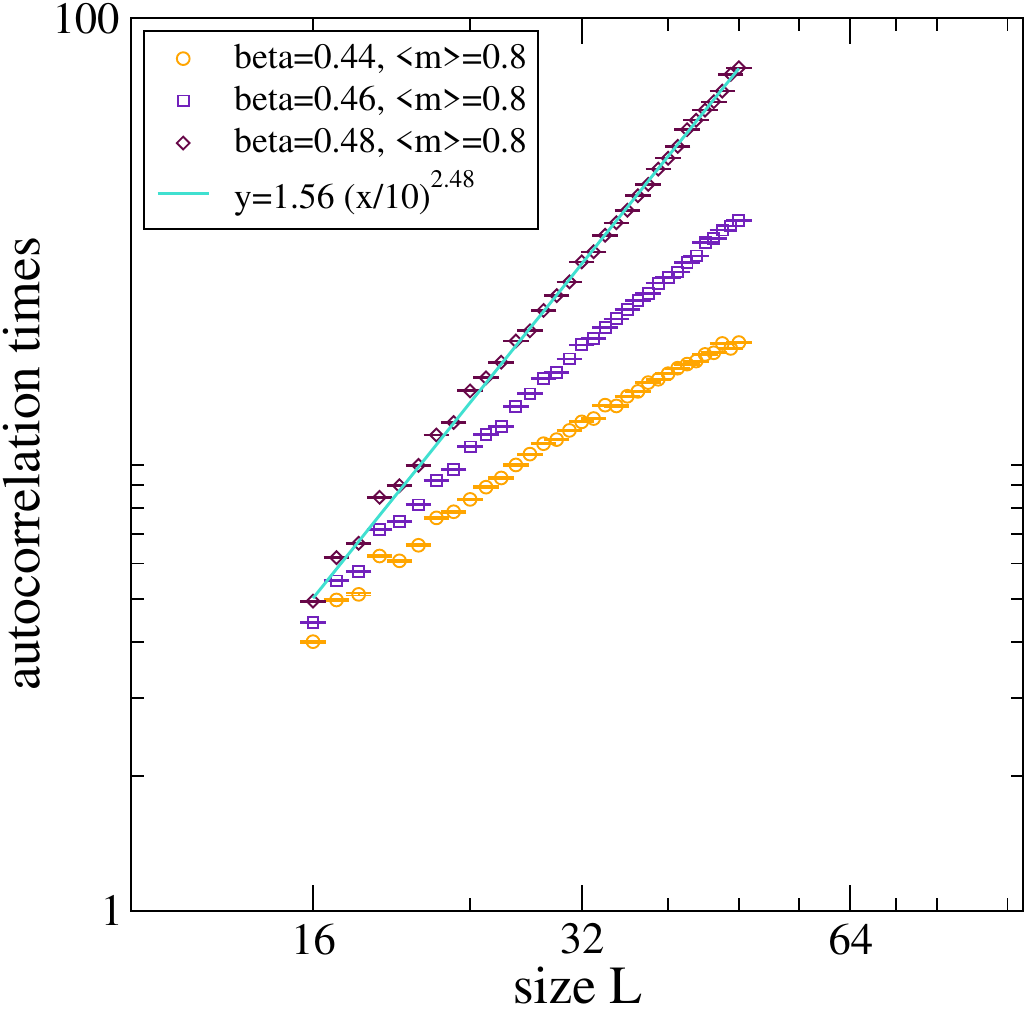}
  \caption{ A comparison of the dependence of autocorrelation time for the observable $M_1$ on lattice size $L$ for the LLR approach for $m_0=0.8$ for several values of $\beta $.} \label{fig:auto_comp3}
\end{figure}

\bigskip
We observe a very small coefficient $b_2$ when compared to the
heatbath simulation where $b_2 \approx 0.28$ at $\beta =0.48$. The
quality are less convincing especially for $\beta = 0.5$. Here,
figure~\ref{fig:8} shows two fits: the exp-powerlaw fit
(\ref{eq:25})and a power-law fit $b_2=0$. Both fits reasonable well
present the data. We are carefully optimistic that any exponential
growth is a quite small rate implying that autocorrelation times are
manageable for realistic lattice sizes. Higher precision data and
perhaps larger lattice sizes are needed to evidence this at a
quantitative level.

\medskip
As detailed above, only the double-expectation values for  $m_0=0$
are afflicted by criticality since, for $m_0 \neq 0$, the $Z_2$
symmetry is explicitly broken by the window function and an
LLR-coefficient $a \neq 0$. Nevertheless, it is important how the
autocorrelation times scale with the lattice size $L$. In the broken
phase, say for $\beta > 0.45$, the marginal distribution for the
magnetisations peak at rather large values $M/V \approx \pm 0.9 $. For
generic observables with a broad domain of support from large portions
of the domain of magnetisation, the dominant contributions from the
LLR integration over the magnetisation  raises from the region around
$M/V \approx \pm 0.9 $. Hence, we studied the volume dependence of the
observable (\ref{eq:51}) as a function of the lattice size $L$ at $m_0 \neq 0$. The
results for $m_0=0.9$ are shown in figure~\ref{fig:auto_comp3} in the
double-log scale in comparison with the $m_0=0$ data. We observe that
auto correlation times are orders of magnitudes smaller than in the
$m_0 = 0$ case. Most importantly however, we find that the increase of
the autocorrelation time with size is at most  polynomial in $L$ and
for $\beta $ values away from its critical value even sub-polynomial.
Log-log scale plot illustrates this in a particularly clear way, 
mapping any power-law dependence to a straight line. Therefore plots of functions 
that grow faster than a power of $L$ appear as bending upwards from a 
straight line, whereas plots of functions with sub-polynomial growth are bending down 
from a straight line.

\medskip
This is an important finding since observables that receive their
dominant contribution from the regions of large magnetisation {\it are
  not} affected by super critical slowing down.

\section{Discussion and conclusions}
\label{sec:discussion}

\begin{figure*}
  \includegraphics[height=8cm]{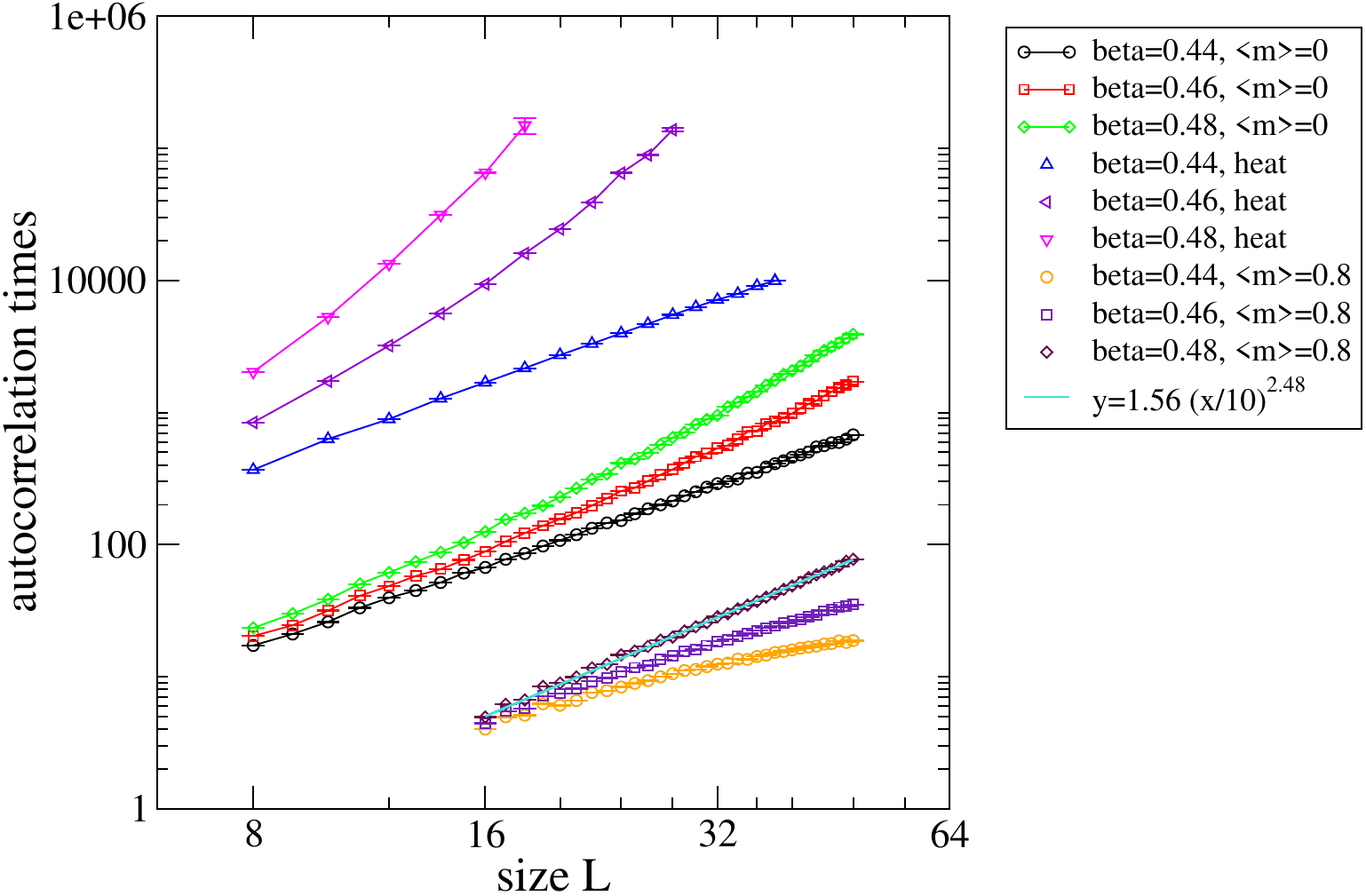}
  \caption{ A comparison of the dependence of autocorrelation time on lattice size $L$ for the conventional heatbath algorithm, where total magnetisation has the longest autocorrelation time, and for the LLR algorithm with magnetisation in the vicinity of$m_0=0$ and $m_0=0.8$, where the Fourier component of magnetisation with lowest nonzero momentum exhibits slowest de-correlation.
  }  \label{fig:auto_comp}
\end{figure*}

Local update algorithms for Markov chains of a given sample size tend
to fail exploring the full configuration space, and hence ergodicity,
for theories in the regime of a spontaneously broken symmetry. In this
regime, the marginal distribution of the order parameter exhibits
several regions of equal stochastic importance but importance sampling
generically selects only one of these regions and fails to transition
between. Consequently, the autocorrelation function rises exponentially
with the system size ({\it super critical slowing down}). A second
question arising is whether the autocorrelation length still rises
polynomial, say at criticality ({\it critical slowing down}). We
adressed both issues in this study. 

\medskip  
Our approach is to decompose the configuration space into the order
parameter as a collective coordinate and the hyperspace orthogonal to
this mode. Wang-Landau techniques (and the LLR method, in particular)
are ideally placed to integrate the slow mode explicitly while the
integration over the hyperspace is done stochastically using MCMC
techniques.  

\medskip 
In this paper, we used a simple two-dimensional Ising
model to demonstrate to explore the performance of the LLR
method. For the Ising model, there are efficient
model-specific cluster algorithms that not only eliminate {\it super
  critical slowing down} but also largely alleviate {\it critical
  slowing down} as witnessed by a small dynamical critical
exponent. Note, however, that cluster algorithms are only available for
very specific models. The present research targets algorithms that
work for a large class of models 'out of the box' without major
fine-tuning. 

\medskip
For the Ising model, the mode that exhibits the longest
autocorrelation time is the global magnetisation, that is, the sum of
all spins. We expect that for all models that are well described by
the Landau theory of phase transitions the global order parameter will
always have the longest autocorrelation time. To confirm this, we also studied the
autocorrelation time for the mode with lowest nonzero momentum $p =
\frac{2 \pi}{L}$, where $L$ is the linear system size. Our approach also
resembles, to some extent, lattice QCD simulations in fixed
topological sectors \cite{Brower:2003yx}. Indeed, global topological
charge is known to be the observable with longest autocorrelation time
in lattice QCD.

\medskip
We found that the LLR algorithm has a potential for solving the
issue of {\it super critical slowing down} for most observables. Only
observables that are sensitive to the marginal distribution around $M
\approx 0$, no matter how small it is, might be affected by critical
slowing down. We only know one such observable: the order-disorder
interface tension. We still see a polynomial rise of the
autocorrelation time with the volume at criticality (and, hence {\it
  critical slowing down}), but we find that at a quantitative level the
autocorrelation time is reduced by orders of magnitude when compared
with that of a heatbath simulation with the same system size (see
figure~\ref{fig:auto_comp}). 

\medskip
As a next step, it would be interesting to check whether explicit
integration over more than one observable using higher-dimensional
generalisation of the LLR algorithm could result in further reduction
of computational time. It is also worth exploring whether the
application of LLR method to fermionic systems could reduce ergodicity
issues related to zeroes of the fermionic determinant. Finally, in a recent
paper \cite{Pawlowski:2022rdn} it was suggested that normalising flows
can eliminate the need to integrate the density of states over $m$ altogether,
thus yielding an even larger speed-up for Monte-Carlo simulations.
It would be interesting to see to what extent normalising flows can further 
reduce the critical slowing down in our situation.

\begin{acknowledgements}
The numerical simulations were undertaken on ARC4, part of the High Performance
Computing facilities at the University of Leeds, UK.
\end{acknowledgements}

\bibliography{lib_crit2}% Produces the bibliography via BibTeX.

\end{document}